\documentclass[iop,apj]{emulateapj}
\slugcomment{The Astrophysical Journal, 753:121 (24pp), 2012 July 10}

\usepackage{xcolor}

\usepackage[bookmarksnumbered,
colorlinks,
linkcolor=blue,
citecolor=black,
filecolor=black,
urlcolor=blue,
breaklinks=true,
]{hyperref}
\shortauthors{Knobel et al.}

\begin{document}

\title{The zCOSMOS\footnotemark[1] 20k group catalog}

\author{C.~Knobel\footnotemark[2],
S.~J.~Lilly\footnotemark[2],
A.~Iovino\footnotemark[11],
K.~Kova\v{c}\footnotemark[2,18],
T.~J.~Bschorr\footnotemark[2],
V.~Presotto\footnotemark[11],
P.~A.~Oesch\footnotemark[2,22],
P.~Kampczyk\footnotemark[2],
%
%enablers
%--------
C.~M.~Carollo\footnotemark[2],
T.~Contini\footnotemark[3,4],
J.-P.~Kneib\footnotemark[5],
O.~Le Fevre\footnotemark[5],
V.~Mainieri\footnotemark[6],
A.~Renzini\footnotemark[7],
M.~Scodeggio\footnotemark[8],
G.~Zamorani\footnotemark[9],
%
%Core team A
%-----------
S.~Bardelli\footnotemark[9],
M.~Bolzonella\footnotemark[9],
A.~Bongiorno\footnotemark[10],
K.~Caputi\footnotemark[2,21],
O.~Cucciati\footnotemark[11],
S.~de la Torre\footnotemark[12],
L.~de Ravel\footnotemark[12],
P.~Franzetti\footnotemark[8],
B.~Garilli\footnotemark[8],
F.~Lamareille\footnotemark[3,4],
J.-F.~Le Borgne\footnotemark[3,4],
V.~Le Brun\footnotemark[5],
C.~Maier\footnotemark[2,20],
M.~Mignoli\footnotemark[9],
R.~Pello\footnotemark[3,4],
Y.~Peng\footnotemark[2],
E.~Perez Montero\footnotemark[3,4,13],
J.~Silverman\footnotemark[14],
M.~Tanaka\footnotemark[14],
L.~Tasca\footnotemark[5],
L.~Tresse\footnotemark[5],
D.~Vergani\footnotemark[9,22],
E.~Zucca\footnotemark[9],
%
%Core team B
%-----------
%
L.~Barnes\footnotemark[2],
R.~Bordoloi\footnotemark[2],
A.~Cappi\footnotemark[9],
A.~Cimatti\footnotemark[15],
G.~Coppa\footnotemark[10],
A.~M.~Koekemoer\footnotemark[16],
C.~L\'opez-Sanjuan\footnotemark[5],
H.~J.~McCracken\footnotemark[17],
M.~Moresco\footnotemark[15],
P.~Nair\footnotemark[9],
L.~Pozzetti\footnotemark[9],
N.~Welikala\footnotemark[19]
}

\footnotetext[1]{European Southern Observatory (ESO), Large Program 175.A-0839}

\affil{$^2$Institute for Astronomy, ETH Zurich, 8093 Zurich, Switzerland\\
$^3$Institut de Recherche en Astrophysique et Plan\'etologie, CNRS, 14 Avenue Edouard Belin, F-31400 Toulouse, France\\
$^4$IRAP, Universit\'e de Toulouse, UPS-OMP, Toulouse, France\\
$^5$Laboratoire d'Astrophysique de Marseille, CNRS/Aix-Marseille Universit\'e, 38 rue Fr\'ed\'eric Joliot-Curie, 13388 Marseille Cedex 13, France\\
$^6$European Southern Observatory, Garching, Germany\\
$^7$INAF-Osservatorio Astronomico di Padova, Vicolo dell'Osservatorio 5, 35122 Padova, Italy\\
$^8$INAF-IASF Milano, Milano, Italy\\
$^9$INAF Osservatorio Astronomico di Bologna, Via Ranzani 1, I-40127 Bologna, Italy\\
$^{10}$Max Planck Institut f\"ur Extraterrestrische Physik, Garching, Germany\\
$^{11}$INAF Osservatorio Astronomico di Brera, Milan, Italy\\
$^{12}$Institute for Astronomy, University of Edinburgh, Royal Observatory, Edinburgh, EH93HJ, UK\\
$^{13}$Instituto de Astrofisica de Andalucia, CSIC, Apartado de Correos 3004, 18080 Granada, Spain\\
$^{14}$Institute for the Physics and Mathematics of the Universe (IPMU), University of Tokyo, Kashiwanoha 5-1-5, Kashiwa, Chiba 277-8568, Japan\\
$^{15}$Dipartimento di Astronomia, Universit\`a degli Studi di Bologna, Bologna, Italy\\
$^{16}$Space Telescope Science Institute, Baltimore, MD 21218, USA\\
$^{17}$Institut d'Astrophysique de Paris, UMR7095 CNRS, Universit\'e Pierre \& Marie Curie, 75014 Paris, France\\
$^{18}$Max Planck Institut f\"ur Astrophysik, Garching, Germany\\
$^{19}$Insitut d'Astrophysique Spatiale, B\^atiment 121, Universit\'e Paris-Sud XI \& CNRS, 91405 Orsay Cedex, France\\
$^{20}$Department of Astronomy, University of Vienna, Tuerkenschanzstrasse 17, 1180 Vienna, Austria\\
$^{21}$Kapteyn Astronomical Institute, University of Groningen, P.O.~Box 800, 9700 AV Groningen, The Netherlands\\
$^{22}$UC Santa Cruz/UCO Lick Observatory, 1156 High Street, Santa Cruz, CA 95064, USA\\
$^{22}$INAF-IASF Bologna, Via P.~Gobetti 101, I-40129, Bologna, Italy
}

\begin{abstract}
\noindent We present an optical group catalog between $0.1 \lesssim z \lesssim 1$ based on 16,500 high-quality spectroscopic redshifts in the completed zCOSMOS-bright survey. The catalog published herein contains 1498 groups in total and 192 groups with more than five observed members. The catalog includes both group properties and the identification of the member galaxies. Based on mock catalogs, the completeness and purity of groups with three and more members should be both about 83\% with respect to all groups that should have been detectable within the survey, and more than 75\% of the groups should exhibit a one-to-one correspondence to the ``real'' groups. Particularly at high redshift, there are apparently more galaxies in groups in the COSMOS field than expected from mock catalogs. We detect clear evidence for the growth of cosmic structure over the last seven billion years in the sense that the fraction of galaxies that are found in groups (in volume-limited samples) increases significantly with cosmic time.  In the second part of the paper, we develop a method for associating galaxies that have only photo-$z$ to our spectroscopically identified  groups. We show that this leads to improved definition of group centers, improved identification of the most massive galaxies in the groups, and improved identification of central and satellite galaxies, where we define the former to be galaxies at the minimum of the gravitational potential wells.  Subsamples of centrals and satellites in the groups can be defined with purities up to 80\%, while a straight binary classification of all group and non-group galaxies into centrals and satellites achieves purities of 85\% and 75\%, respectively, for the spectroscopic sample.\\[2mm]
\noindent \emph{Online-only material:} machine-readable tables
\end{abstract}

%\keywords{catalogs -- cosmology: observations -- galaxies: groups and clusters: general -- galaxies: evolution -- large-scale structure of universe -- methods: data analysis}

%\emph{Key words:} catalogs -- cosmology: observations -- galaxies: groups and clusters: general -- galaxies: evolution -- large-scale structure of Universe -- methods: data analysis\\

%\emph{Online-only material:} machine-readable tables\\

\section{Introduction}
\setcounter{footnote}{21}
Galaxy groups are gravitationally bound systems that contain multiple galaxies inhabiting the same dark matter (DM) halo. They are of interest for two main reasons. First, by regarding them as DM halos they can serve as cosmological probes. The number density and clustering of groups for a given halo mass and cosmic epoch depend on the underlying cosmology. Second, galaxy groups constitute an environment for galaxies which is special compared with the general field.  The enhanced proximity of other galaxies and the presence of an intragroup medium may produce distinct evolutionary processes in groups such as enhanced merging rates \citep{spitzer1951}, galaxy harassment \citep{moore1996}, ram pressure stripping \citep{gunn1972}, or strangulation \citep{balogh2000} which may be significant for the general evolution of galaxies, and particularly the environmental differentiation of the galaxy population \citep[e.g.,][]{weinmann2006,gerke2007,iovino2010,kovac2010,peng2010}.  A difference between central and satellite galaxies in groups is now an established part of our view of galaxy evolution \citep[e.g.,][]{vandenbosch2008,skibba2009a,pasquali2010,skibba2011,peng2011}. A key requirement for both areas is the availability of large, high quality group catalogs.

There are several desired properties for a group catalog.  Purity and completeness are two often conflicting requirements---completeness is the fraction of real groups that are recovered, while purity reflects the reality of the claimed groups.  Once the groups are identified, one can further define purity and completeness for the membership of individual galaxies in these groups.   The optimization between completeness and purity will often depend on the application: high purity catalogs covering a large redshift range enable studies of galaxy evolution in different environments over cosmic time. On the other hand, having complete catalogs which trace the numbers of real groups and provide reliable mass estimates for individual groups is important for cosmological studies.  The estimation of reliable masses for individual groups in turn requires a high degree of one-to-one correspondences between reconstructed and real groups.  Precise estimates for the group centers is needed for stacking analyses of X-ray properties or detection of the weak lensing signal, while studying the differences between ``central'' and ``satellite'' galaxies requires complete group populations down to a given flux limit since otherwise the central galaxy cannot be reliably identified.

In this paper we present a new group catalog produced with the zCOSMOS-bright survey \citep{lilly2007}, which now contains about 16,500 high quality spectroscopic galaxies with $I_{\rm AB} \leq 22.5$ in the redshift range $0.1 \lesssim z \lesssim 1.2$ (the ``20k sample''). zCOSMOS-bright covers the $\sim \! 1.7$ deg$^2$ of the COSMOS field \citep{scoville2007} which was fully observed by the \emph{Hubble Space Telescope} \citep{scoville2007a,koekemoer2007} down to $I_{\rm AB} < 28$ ($5\sigma$) and followed up in more than 30 bands by several telescopes from radio to X-ray wavelengths \citep{capak2007}. This unique combination of observational data on a single field makes the COSMOS field very suitable for studying the properties of groups as a function of redshift and the evolution of galaxies in different environment. The large numbers of wavelength bands also allows the production of high quality photometric redshifts (``photo-$z$'') with an accuracy of $\delta z \sim 0.01 (1 + z)$ \citep[e.g.,][]{ilbert2009} for the brighter galaxies, allowing the possibility of using these to supplement the spectroscopic redshifts and assign, at least probabilistically, group membership to these galaxies.

The first major data release of zCOSMOS entailed about 8,500 spectroscopic galaxy redshifts (the ``10k sample'', \citealt{lilly2009}) and was used to produce a first optical group catalog in the redshift range $0.1 \lesssim z \lesssim 1$ (\citealt{knobel2009}, ``K09''). In that paper we discussed in detail the group-finding methods and basic properties of the ``10k group catalog''. We adopted two group-finding algorithms, friends-of-friends (FOF) and a Voronoi-Delaunay method (VDM), and compared their performances on simulated mock galaxy samples.  We introduced a ``multi-run scheme'' in which we successively used different group-finding parameters, optimized for different richness groups, where by richness we always refer to the number $N$ of observed spectroscopic members.  By initially tuning the parameters to detect only the richest groups, and then ignoring the subsequent fragmentation of these into smaller groups when the parameters were tuned to smaller scales, we could improve the statistics of the catalog in terms of completeness and purity over a wide range of scales, minimizing the effects of fragmentation and overmerging (see K09 for a discussion).  The FOF catalog was used as the basic 10k group catalog while the VDM catalog was used to produce subcatalogs with further enhanced purities.  The basic 10k catalog contained 802 groups in total and 102 groups with more than five members.

The group catalog presented in this paper is created in a similar way to that in K09 from the larger sample that is now available. However, since it now contains groups extending up to 30 members, we had to slightly extend the methods to guarantee its high quality over this wider range of richness. 

In contrast to the zCOSMOS 10k sample whose completeness was only about 30\% and for which it would have made little sense to use information from photo-$z$, the completeness of the 20k sample now exceeds 50\% and the photo-$z$ objects become a minority.  Thus it becomes attractive to try to associate these remaining photo-$z$ objects to the spectroscopically identified groups, so that an idea of group membership can be obtained for all galaxies down to the magnitude limit of the survey.  This is useful for many scientific goals.  We therefore develop a method for incorporating the photo-$z$ galaxies into the spectroscopic group population by assigning to each photo-$z$ galaxy a probability that it is a member of a given group. This probability is based on the projected spatial distance of the galaxy from the group center and its photo-$z$ relative to the redshift of the group, calibrated against mock catalogs. Including the photo-$z$ galaxies enables improved estimates of the location of the group center, and improved identification of the most massive galaxy in the group, and of the galaxy lying at the center of the potential well, which we define as the central galaxy.  For the latter two cases, we can construct various samples which represent trades between completeness and purity.  As a result, we also look into how well we can apply a binary central-satellite classification to all galaxies in the sample, including those not associated with groups.

With the final 20k sample we produce a group catalog containing almost 1500 groups in the redshift range $0.1 \lesssim z \lesssim 1$.  Other major group catalogs at redshift $z \gtrsim 0.3$ are the one from the DEEP2 survey \citep{davis2003} containing $\sim\! 2400$ groups \citep{gerke2005,gerke2012} in the redshift range $0.7 \lesssim z \lesssim 1.4$ and the one from VVDS \citep{lefevre2005} containing $\sim\! 300$ groups in the redshift range $0.2 \leq z \leq 1$ \citep{cucciati2010}, so the new zCOSMOS catalog is one of the largest published group catalogs at high redshift (and the largest on a contiguous field) and features very good statistics compared to the other group catalogs at high redshift in the literature. A special feature of our group catalog is the availability of group centers that are based on a sophisticated approach and the possibility to produce high-purity samples of central and satellite galaxies.

This paper is organized as follows. In Section~\ref{sec:data}, we describe the observational and mock data used for our work. In Section~\ref{ch:spectroscopic_group_finding_method}, we describe the method of group identification and the statistical results obtained using the mock catalogs. We then give a detailed description of the final zCOSMOS spectroscopic group catalog in Section~\ref{sec:spectroscopic_group_catalog} and perform some comparisons with the mock catalogs. In the second part of the paper, we first develop, in Section~\ref{ch:photometric_group_population_method}, the method for associating photo-$z$ galaxies to the spectroscopically identified groups. We then discuss in Section~\ref{sec:applications} how this can lead to improved definitions of the corrected richness of the groups, of the most massive galaxies, of the spatial centers, and of the central galaxies, defined as those at the bottom of the potential well. The properties of the centrals and satellites will be explored in two further papers in preparation (C.~Knobel et al., in preparation; K.~Kova\v{c} et al., in preparation). In Section~\ref{sec:discussion} we finally comment on the general difficulties in producing high quality group catalogs and in Section~\ref{sec:conclusions} we conclude the paper.

In the paper we will frequently make comparison with the set of 24 mock catalogs, which are 24 different realizations of a single model universe. When we apply a general algorithm to the mock catalogs, the scatter among the 24 returned values represents the minimum uncertainty that can be expected when we apply the algorithm to the actual data, due to issues such as cosmic variance. We refer to this as the standard deviation of the relevant parameter among the mock catalogs.  It is this scatter which is appropriate when we wish to consider whether the real data are or are not consistent with the model universe of the mock catalogs. The best estimate of the overall performance of the algorithm in question obviously comes from the average of all 24 mock catalogs.  The uncertainty in this estimate is given by the standard deviation above divided by $\sqrt{24}$. We will refer to this as the standard deviation of the mean.

Where necessary, a concordance cosmology with $H_0 = 70\ \rm{km\; s^{-1}\; \rm{Mpc}^{-1}}$, $\Omega_{\rm m} = 0.25$, and $\Omega_\Lambda = 0.75$ is applied. All magnitudes are quoted in the AB system. We use the term ``dex'' to express the antilogarithm, i.e., 0.1 dex corresponds to a factor $10^{0.1} \simeq 1.259$.

\section{Data}
 
In this section we describe the data that have been used for this paper. First, we give an overview of the zCOSMOS survey from which the spectroscopic redshifts are taken, then we describe the derivation of the photometric redshifts, masses, and absolute magnitudes using the photometry of the COSMOS survey, and finally we describe the construction of realistic mock galaxy samples.
 
\subsection{The zCOSMOS survey} \label{sec:data}
 
zCOSMOS \nocite{lilly2007,lilly2009}(Lilly et al.~2007, 2009; S.~J.~Lilly et al. 2012, in preparation)
%\citep[][S.~L.~Lilly et al. 2012, in preparation]{lilly2007,lilly2009}
is a deep spectroscopic galaxy survey on the 1.7 deg$^2$ of the COSMOS field \citep{scoville2007} which utilized about 600 hours of ESO VLT service mode. It is divided up into two parts, ``zCOSMOS-bright'' and ``zCOSMOS-deep''. The former covers mainly the redshift range $0.1 \lesssim z \lesssim 1.2$ and almost the entire COSMOS field, while the latter aims to cover the redshift range $1.5  \lesssim z \lesssim 3$ on the central $\sim \! 1$ deg$^2$ of the COSMOS field.

The current work is entirely based on zCOSMOS-bright, which is now complete and contains spectra for about 20,000 objects taken using the VIMOS spectrograph \citep{lefevre2003a} with a medium-resolution grism. The target catalog consisted basically of all objects within the magnitude range $15 \leq I_{\rm AB} \leq 22.5$. Suspected stars were excluded. The slits were assigned to the targets such that for each mask the number of slit assignments on each of the four VIMOS quadrants was maximized---except for some X-ray and radio objects which were observed at high priority. Since there were two masks per pointing and the pointings were overlapping with centers differing by the size of a quadrant, there were finally eight passes for the central field, four at the borders, and two at the corners.

About 2\% of all spectra come from ``secondary'' objects, i.e., objects that were potential targets which serendipitously ended up in slits targeted at other galaxies. They are not only very helpful for estimating the accuracy and verification rate of redshifts, but also compensate for the bias against close pairs due to slit constraints \citep{deravel2011,kampczyk2011}. After removing less reliable redshifts (i.e., confidence classes 0, 1.1, 2.1 and 9.1; see \citealt{lilly2009}) and spectroscopic stars, we end up with a high quality redshift galaxy sample containing 16,776 objects within the area $149.47^\circ \lesssim \alpha \lesssim 150.77^\circ$ and $1.62^\circ \lesssim \delta \lesssim 2.83^\circ$. From multiply observed objects the spectral verification rate for this sample is about 99\% and the redshift accuracy about 100 km s$^{-1}$ which is sufficient to probe the cosmic group environment. The remaining objects and all those not observed spectroscopically have photo-$z$ available. Henceforth we will refer to this sample of secure redshifts as the ``20k sample''.

\begin{figure}
	\centering
	\includegraphics[width=0.46\textwidth]{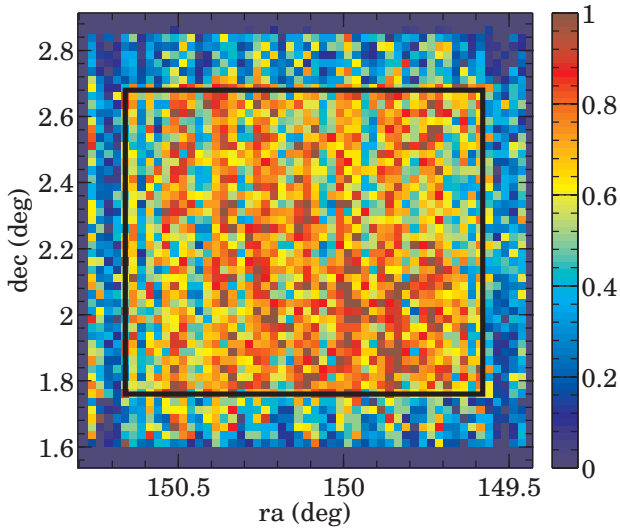}
	\caption{Spatial sampling rate (SSR) of the zCOSMOS 20k sample. The color bar indicates the SSR, which is computed in pixels of 1.5 arcmin. The black rectangle shows the central region for the 20k sample.\label{fig:mask}}
\end{figure}

The spatial sampling rate (SSR), i.e., the fraction of objects of the magnitude-limited target catalog whose spectra were observed, is a function of $(\alpha, \delta)$ and is shown in Figure \ref{fig:mask}.
According to the design of zCOSMOS there is a central region ($\alpha = 150.12\pm 0.54^\circ$ and $\delta = 2.22\pm 0.46^\circ$, see the black rectangle) with a substantial higher SSR than at the borders. Even in the central region, the SSR is not completely uniform, exhibiting some stripes due to the placement of slits in the masks. The redshift success rate (RSR) is the fraction of observed spectra that have yielded a reliable redshift.  The RSR is mostly a function of apparent magnitude and redshift of the galaxies and only weakly dependent on color (see Figs.~2 and 3 of \citealt{lilly2009}).

Approximately, the SSR and RSR can be assumed to be uncorrelated so that by multiplying them we obtain for each galaxy the completeness in respect to an ideal magnitude-limited survey. The full zCOSMOS area has an average completeness of 48\%, while for the central region it rises to 56\%. For some applications it is useful to restrict the area of the survey to the central region where the sampling rate is highest.  It should be noted that the redshift distribution of galaxies in the COSMOS field shows two prominent features at redshifts $\sim\! 0.35$ and $\sim\! 0.7$ (cf.~Fig.~1 of K09).

\subsection{Photometric redshifts} \label{sec:photo}

Photometric redshifts (photo-$z$), masses, and absolute magnitudes were derived from spectral energy distribution (SED) fitting using \emph{ZEBRA}+ \citep[][]{oesch2010},  which is an extension of \emph{ZEBRA} \citep{feldmann2006}, to allow for the derivation of physical properties of the galaxies using stellar population synthesis models.

The photo-$z$ were derived from a fit of empirical templates to 26 photometric bands from $u^\ast$ (CFHT) to \emph{Spitzer} IRAC4.8 including 12 broad-band, 12 intermediate band, and 2 narrow-band filters. The empirical template set was based on \cite{bruzual2003} models, to which emission lines were added, before running the template correction module of \emph{ZEBRA} based on a random subsample of zCOSMOS spectroscopic redshifts. For the few hundred \emph{XMM-Newton} X-ray sources the photo-$z$ provided by \cite{salvato2009} were taken \citep[published in][]{brusa2010}.

The stellar masses were subsequently derived from standard \cite{bruzual2003} models with an initial stellar mass function of \cite{chabrier2003} and dust extinction according to \cite{calzetti2000}. Due to the absence of emission lines in the model SEDs, only the broad-band photometry was used for the SED fit, where the redshift was fixed at the spec-$z$ of the galaxy, if available, or otherwise at the adopted photo-$z$.

In order to increase the fidelity of our photo-$z$ sample we excluded 5\% objects by applying a cut in the resulting $\chi^2$ from the SED fit and required that for each object at least nine broad band filters were available. Comparison with the spectroscopic control sample yielded a photo-$z$ error of about $0.01(1+z)$ and a catastrophic failure rate of 2-3\% where a catastrophic failure is defined by $|z_{\rm spec}-z_{\rm phot}|>0.04(1+z)$. (The subsample that was excluded had catastrophic failure rate of $\sim 60\%$.) We compared our stellar masses to those derived using \emph{Hyperzmass} \citep[see][]{bolzonella2010} which yielded an uncertainty in stellar mass of about 0.2 dex. Note that the stellar masses were derived without considering mass return in the sense that ``stellar mass'' is simply the integral of the star formation rate, since this is more useful for most purposes. These masses are typically 0.2 dex larger than when considering mass return.

\subsection{Mock catalogs}\label{sec:mocks}

The mock catalogs that are used for tuning the group-finding parameters and for comparing our results with cosmological simulations are adapted from the COSMOS mock light cones \citep{kitzbichler2007} which are based on the Millennium DM $N$-body simulation \citep{springel2005} run with the cosmological parameters $\Omega_{\rm m} = 0.25$, $\Omega_\Lambda = 0.75$, $\Omega_{\rm b} = 0.045$, $h = 0.73$, $n = 1$, and $\sigma_8 = 0.9$. The semi-analytic recipes for populating the DM halos with galaxies are that of \cite{croton2006} as updated by \cite{delucia2007}. There are 24 independent mock catalogs, each covering an area of $1.4\ \rm{deg} \times 1.4\ \rm{deg}$ with an apparent magnitude limit of $r \leq 26$ and a redshift range of $z \lesssim 7$.

The mock catalogs were adjusted to resemble as closely as possible the actual 20k sample. For details we refer to K09.  After applying a magnitude cut the mean number of galaxies in the mock catalogs (averaged over all 24 fields) are slightly different from the number of galaxies in the zCOSMOS target catalog (a 1$\sigma$--2$\sigma$ effect).  Since the density of galaxies is important for tuning the group-finding parameters, we applied a small adjustment, uniform across all mock catalogs and smoothly varying in redshift, to the magnitude limit for the mock catalogs so as to match the correct (smoothed) number of galaxies with redshift. This intervention has, however, only a very small effect on the analysis in this paper and we usually checked that our results did not depend sensitively on this alteration. We then applied the SSR and RSR to the mock catalogs by randomly removing galaxies from the magnitude-limited mock sample and implemented a Gaussian redshift measurement error of $\delta z = 100(1+z)/c\ \rm{km\ s^{-1}}$.

For the second part of the paper, we extend the spectroscopic 20k mock samples by adding simulated photo-$z$ galaxies so that the spec-$z$ and photo-$z$ mock samples add up to the $I_{\rm AB} \leq 22.5$ complete samples for each mock catalog. That is, each galaxy brighter than the flux limit that is not part of the spec-$z$ mock sample was assigned a photometric redshift by perturbing its original redshift by an amount drawn from a Gaussian distribution with standard deviation $\delta z = 0.01(1+z)$. We also perturbed the stellar masses of all galaxies, spec-$z$ as well as photo-$z$, by adding a Gaussian random number with standard deviation of 0.2 to $\log(M/M_\odot)$ to mimic the stellar mass uncertainty of 0.2 dex of the actual data.

\section{Group-finding method}\label{ch:spectroscopic_group_finding_method}

In this section, we describe the method of group identification and provide the resulting group catalog statistics as obtained with the mock catalogs. We will slightly modify the methods presented in K09 to optimize them for the 20k sample. A novelty of the 20k group catalog is the existence of a larger number of relatively rich groups with $N > 12$, so that the optimization strategy has to be adapted to yield stable statistics for these higher richness classes as well. The application to the zCOSMOS 20k sample is presented in the next section.

\subsection{Definitions}\label{sec:definitions}

We will mainly follow the terminology and statistics introduced in Section~3.2 of K09 which shall be briefly summarized in the following. For details we refer to K09.

\begin{figure}
	\centering
	\includegraphics[width=0.46\textwidth]{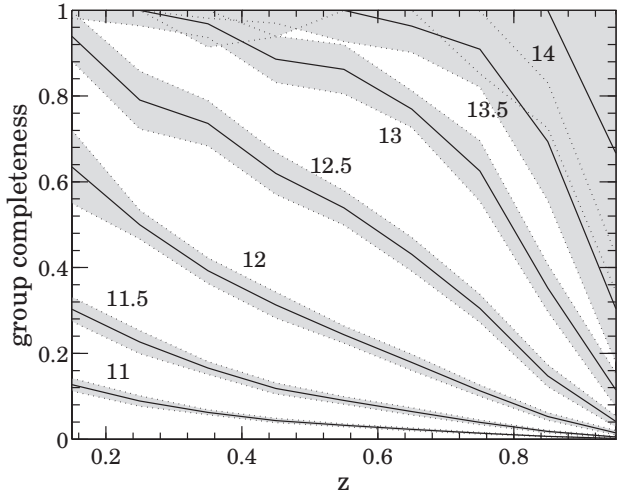}
	\caption{Fraction of detectable halos in the zCOSMOS 20k mock samples, as a function of redshift, where detectable corresponds to having at least two members with spectroscopic redshifts above $I_{\rm AB} = 22.5$ after the spacial sampling and spectroscopic success rate are applied. The lines (from bottom to top) correspond to groups more massive than 11, 11.5, 12, 12.5, 13, 13.5, and 14, respectively, in units of $\log(M/M_\odot)$. The shaded area is the standard deviation among the 24 mock catalogs.\label{fig2:halo_completeness}}
\end{figure}

A group is defined as the set of galaxies occupying the same DM halo.\footnote{Since the groupfinder are calibrated using the mock catalogs, the definition of a DM halo used in this paper corresponds to the operational definition of a DM halo in the Millennium simulation. That is, a DM halo is a friends-of-friends group of DM particles with a linking length of $b = 0.2$. These groups ideally correspond to structures with a mean overdensity of roughly 200.} In the mock catalogs we know exactly which galaxies are in which groups and we denote the corresponding sets of galaxies as the ``real groups''. On the other hand, the set of groups obtained by running a groupfinder on actual or mock data are called ``reconstructed groups''. The aim of group-finding is to tune the parameters of the groupfinder so that the resulting catalog of reconstructed groups approaches as closely as possible the catalog of real groups, as measured by certain statistics. It should be stressed that the ``real'' groups correspond to those DM halos which would be ``detectable'' (i.e., which host at least two galaxies with spectroscopic redshift measurements) in a galaxy survey with the same characteristics as zCOSMOS. Figure \ref{fig2:halo_completeness} shows the fraction of these detectable DM halos as compared to the overall sample of all DM halos.
Note that more than $90\%$ of DM halos of mass $>\! 10^{13.5}\ M_\odot$ are detectable up to a redshift of $z \simeq 0.8$, while, for groups more massive than $10^{12.5}\ M_\odot$, the completeness decreases linearly with redshift from $\sim\! 90\%$ at $z \simeq 0$ down to $\sim\! 10 \%$ at $z \simeq 0.9$.

With the concepts of the ``real'' and ``reconstructed'' groups we can define the ``completenesses'' and ``purities'' of samples of reconstructed groups by associating the real groups to reconstructed groups and vice versa.  A real (reconstructed) group is associated to a reconstructed (real) group if the former contains more than 50\% of the members of the latter. All such associations are called ``one-way-match'' (1WM). If the association is mutual then we call it also a ``two-way-match'' (2WM; see Fig.~3 of K09 for illustration).  2WM are thus 1WM.

In K09 we demonstrated that the statistics of the group catalog can strongly depend on the richness $N$, which is the number of observed spectroscopic members for a given group, and we introduced the multi-run scheme to overcome this.   To check this aspect of our catalog, we investigate the statistics as a function of $N$ in what follows. It should be noted that $N$ will be biased with respect to redshift, since it refers to galaxies above the survey flux limit, and it is also affected by the local sampling rate. Hence, the richness $N$ is a parameter that describes the identification of a group, and the amount of information about it, rather than the actual number of galaxies that reside in it. To obtain an estimate of the actual number of members, unbiased with respect to redshift, the corrected richness (see Sects.~\ref{sec:physical_properties} and \ref{sec:corrected_richness}) should be used defined in terms of a volume-limited galaxy sample.

The one-way completeness $c_1(N)$ is then defined as the number of 1WM of real groups of richness $N$ to reconstructed groups of any richness divided by the number of real groups of richness $N$. Note that in K09 we defined these quantities in a cumulative way, i.e., always for $\geq\!N$, here we define them as functions of $N$ only. The two-way completeness $c_2(N)$ is similarly defined by considering 2WM instead of 1WM. Similarly, the one-way purity $p_1(N)$ is defined by the number of 1WM of reconstructed groups of richness $N$ to real groups of any richness normalized by the number of reconstructed groups of richness $N$, and the two-way completeness $p_2(N)$ is obtained by exchanging 1WM with 2WM.  While these statistics are made on a group-by-group basis, there are analogous statistics, referring to individual galaxy memberships in groups, which are the galaxy success rate $S_{\rm gal}(N)$ in correctly assigning group membership to galaxies, and the interloper fraction $f_{\rm I}(N)$ which gives the fraction of non-group galaxies that are incorrectly assigned to groups.

In addition to these statistics we also introduced in K09 the figures of merit $g_1$ and $g_2$:
\begin{eqnarray}
g_1(N) &=& \sqrt{(1-c_1(N))^2 + (1-p_1(N))^2} \label{eq:g1}\\
g_2(N) &=& \frac{c_2(N)}{c_1(N)}\;\frac{p_2(N)}{p_1(N)}\:.
\end{eqnarray}
They are defined such that they are numbers in the interval between 0 and 1. $g_1(N)$ is a measure of the balance (or trade-off) between 1WM completeness and purity and $g_2(N)$ is a measure of the balance between fragmentation and overmerging of reconstructed groups. For a good group catalog $g_1$ should be close to zero and $g_2$ close to one for all ranges of richnesses. In this paper, we introduce another figure of merit
\begin{equation}
\tilde g_1(N) = \sqrt{(1-c_2(N))^2 + (1-p_2(N))^2} \label{eq:g1_tilde}
\end{equation}
which is similar to $g_1$ except that all 1WM statistics are replaced by their 2WM statistic counterparts.

We remind readers that these statistics compare the reconstructed group catalog to the real group catalog, i.e., to the groups that are in principle detectable within zCOSMOS.

\subsection{Optimization strategy}\label{sec:optimization_strategy}

The basic group-finding algorithms we apply are the FOF and VDM algorithms that were described in Section~3.1 of K09. The main task is to optimize the group-finding parameters such that the resulting catalog exhibits the best possible statistics. For the 10k sample the group-finding strategy was mainly driven by minimizing $g_1(N)$ for several richness classes. However, since $g_1(N)$ is only based on 1WM statistics, it does not account for fragmentation or overmerging in the resulting catalog. Thus, if optimized for $g_1(N)$ the resulting catalog might contain, unnecessarily, many such overmerged or over-fragmented groups which will exhibit very good one-way statistics but very poor two-way statistics. A reconstructed group that is fragmented or overmerged will fail to tell us anything about the true nature of the group such as its mass, richness, or radius. It will only tell us if a certain galaxy is a group galaxy or not. Therefore, the number of such groups should be kept as low as possible. This is why we decided in the present work to optimize the parameters for the modified $\tilde g_1(N)$ instead of $g_1(N)$.

Optimizing the single-richness runs in respect to $\tilde{g}_1$ instead of $g_1$ will, of course, yield slightly worse $g_1$ values for the single runs. This, however, does not have to be true for the $g_1(N)$ statistics of the global multi-run catalog. The combination of several single runs with inferior $g_1$ statistics can lead to a multi-run catalog with slightly superior $g_1$ for small $N$ than the multi-run catalog of the single $g_1$-optimized single runs. This seeming paradox is resolved by noting that, in a multi-run scheme, the single runs can interfere in a complicated nontrivial way. For instance, if the first run being optimized for large groups aims to produce a very complete catalog, it will lead to some overmerging of some parts of small groups, which cannot then be detected in later runs. As a result, the first run can already spoil the $g_1$ statistics of the small groups.

How can the parameters of the single runs be optimized in order to produce an optimal multi-run catalog? This is probably the most difficult part in the overall group-finding procedure and, unfortunately, there is no general prescription in order to produce ``the'' unique optimal multi-run catalog. In principle, one would have to analyze the statistics of the multi-run catalog for all possible parameter combinations of the single runs. This would not only be computationally very expensive, but would also require a distinct single figure of merit for characterizing a whole catalog.\footnote{It is unlikely that such a single optimal figure of merit exists. For instance, the optimal catalog in respect to $\tilde{g}_1$ over the whole range of group sizes is not necessarily also the optimal catalog in respect to the produced number of reconstructed groups $N_{\rm rec}$, since we found that almost equally good catalogs in respect to $\tilde{g}_1$ can exhibit substantial differences in $N_{\rm rec}$.}

Thus, a manageable way of producing an optimized multi-run catalog is to first produce a couple of optimized single runs and then try different combinations always keeping an eye on $\tilde{g}_1(N)$ and the number of reconstructed groups $N_{\rm rec}(N)$. As a guideline the parameters of the single runs which are to be combined to the multi-run should not exhibit any large discontinuities as a function of richness. That is, the parameters of the multi-run should be slowly varying as we move down to smaller and smaller groups.

While this approach works pretty well for FOF, it is less convenient for the VDM parameters because their effect on the final catalog statistics is much harder to anticipate intuitively and it is even harder to anticipate the effect of different combinations of single runs. The final parameter sets for the FOF and VDM 20k multi-run catalogs are given in Table \ref{tab2:parameter-sets-FOF-20k} and \ref{tab2:parameter-sets-VDM-20k}, respectively. Note that the justification for these particular parameter-sets is based only on the extremely good statistics of the final product (see Sect.~\ref{sec:statistics}) and not by any rigorous optimization procedure.  Moreover, we have also checked that the application of these group-finding parameters on the actual data yield consistent behavior between the actual data and the mock catalogs, e.g., in the number of 1WM between FOF and VDM (cf.~Fig.~7 of K09).
\begin{deluxetable}{ccrrclll}
%\tablewidth{0.7\textwidth}
\tablewidth{0pt}
\tablecaption{Multi-run parameter sets for FOF}
\tablehead{
	\colhead{Step} &
	\colhead{} &
  \colhead{$N_{\rm min}$} &
  \colhead{$N_{\rm max}$} &
  \colhead{} &
  \colhead{$b$} &
  \colhead{$l_{\rm max}$} &
  \colhead{$R$} \\
  \colhead{} &
	\colhead{} &
  \colhead{} &
  \colhead{} &
  \colhead{} &
  \colhead{} &
  \colhead{(Mpc)\tablenotemark{a}} &
  \colhead{} 
  }
\startdata

1 && 11 & 500 && 0.1    & 0.375 & 18.5  \\
2 && 7  & 10  && 0.095  & 0.38  & 14.5  \\
3 && 6  & 6   && 0.09   & 0.35  & 16    \\
4 && 5  & 5   && 0.085  & 0.375 & 13.5  \\
5 && 4  & 4   && 0.075  & 0.3   & 19.5  \\
6 && 3  & 3   && 0.09   & 0.275 & 18.5  \\
7 && 2  & 2   && 0.06   & 0.225 & 16.5  

\enddata

\tablenotetext{a}{Physical length.}

\label{tab2:parameter-sets-FOF-20k}
\end{deluxetable}

\begin{deluxetable*}{ccrcrcrcr}
\setlength{\tabcolsep}{1mm}
\tablewidth{0pt}
\tablecaption{Multi-run parameter sets for VDM}
\tablehead{
	\colhead{Step} &
  \colhead{$N_{\rm min}$} &
  \colhead{$N_{\rm max}$} &
  \colhead{$R_{\rm I}$} &
  \colhead{$L_{\rm I}$} &
  \colhead{$R_{\rm II}$} &
  \colhead{$L_{\rm II}$} &
	\colhead{$r$} &
  \colhead{$l$} \\
  \colhead{} &
  \colhead{} &
  \colhead{} &
  \colhead{(Mpc)} &
  \colhead{(Mpc)} &  
  \colhead{(Mpc)} &
  \colhead{(Mpc)} & 
  \colhead{(Mpc)} &
  \colhead{(Mpc)} 
  }
\startdata

1 & 9 & 500 & 0.7 & 12 & 0.7 & 10 & 0.7 & 10 \\
2 & 5 & 8   & 0.7 & 12 & 0.4 & 8  & 0.5 &  8 \\
3 & 2 & 4   & 0.4 &  8 & 0.4 & 8  & 0.5 &  7

\enddata

\tablecomments{All units of lengths are comoving.}

\label{tab2:parameter-sets-VDM-20k}
\end{deluxetable*}

Looking at Figure \ref{fig:mask} one might be tempted to introduce a spatially variable linking length to account for the variations in the projected density of galaxies caused by variations in the SSR. We carried out tests of this by implementing, for example, sinusoidally varying linking lengths along the right ascension axis that produced slightly larger values in underdense strips than in the overdense strips in Figure \ref{fig:mask}. Interestingly, our optimization scheme preferred a non-varying linking length.   The reason for this is that, except at low redshifts $z \lesssim 0.3$, the FOF linking length $l_\perp$ is set by the maximum linking length $L_{\rm max}$ (see K09), which is introduced to be of the order of the expected physical size of the DM halos, which is thus independent of the local galaxy density. This is also demonstrated if we allow a general functional form for the redshift dependence of the linking length $l_\perp(z)$. The preferred redshift dependence, in terms of optimizing the statistics of the group catalog, is a linking length that is basically constant in physical space, even though the density of galaxies drastically decreases with redshift.   This is also seen in the fact that the statistics of the group catalog are very similar whether we consider the full COSMOS field or only the central region (see Tab.~\ref{tab2:catalog_properties}).

As discussed and implemented in K09, a much more important effect is that the optimal linking length $l_\perp$ depends on richness.  This motivated our multi-run scheme. As shown in Table \ref{tab2:parameter-sets-FOF-20k} the linking lengths for the seven different runs in this scheme differ by up to 50\%.

\subsection{Subcatalogs}\label{sec:sub_catalogs}

As in K09, we take the FOF multi-run group catalog to be the main group catalog and use the VDM multi-run catalog to define the galaxy purity parameter, GAP$_i$, for $i \in \{1,2\}$ as follows: if an FOF group galaxy is also in a VDM group such that there is a 1WM between the FOF and the VDM group, the GAP$_1$ of this galaxy is set to 1, and to 0 otherwise. Similarly, if there is a 2WM between these groups, then the GAP$_2$ is 1, and 0 otherwise.

This concept can be generalized to a group as a whole by computing the fraction of members of a given group that have a GAP $\neq 0$. We define the group purity parameter, GRP$_i$, $i = \{1,2\}$ of a group to be the fraction of galaxies in that group that have GAP$_i = 1$. By selecting those groups with a GRP$_i$ larger than some threshold, we generate subcatalogs of the original FOF group catalog with higher purity, as shown in the next paragraph. The subcatalog consisting of all groups with GRP$_i > 0$ excludes groups that are only detected in FOF.  We call this the GRP$_i$ subcatalog.\footnote{As an aside, the GRP$_i$ catalogs are similar but not identical to what we called the $i$WM, $i = \{1,2\}$ subcatalogs in K09. The $i$WM catalogs contained not only a subsample of groups of the basic FOF catalog, but also a subsample of the members of each group so that the richness of a group of the $i$WM catalog was in general not the same like that of the corresponding FOF group. For the groups of the GRP$_i$ catalogs, the richness is always the same. In this paper, we will never use the term $i$WM in the meaning of subcatalogs as in K09, but only to indicate the relation between reconstructed and real groups.}

It turns out that the statistics of the basic FOF catalog and its GRP$_1$ subcatalog are very similar. Consequently we omit the latter in the following, including instead just the GRP$_2$ subcatalog.

\subsection{Catalog statistics for the mock catalogs}\label{sec:statistics}

\begin{figure}
	\centering
	\includegraphics[width=0.45\textwidth]{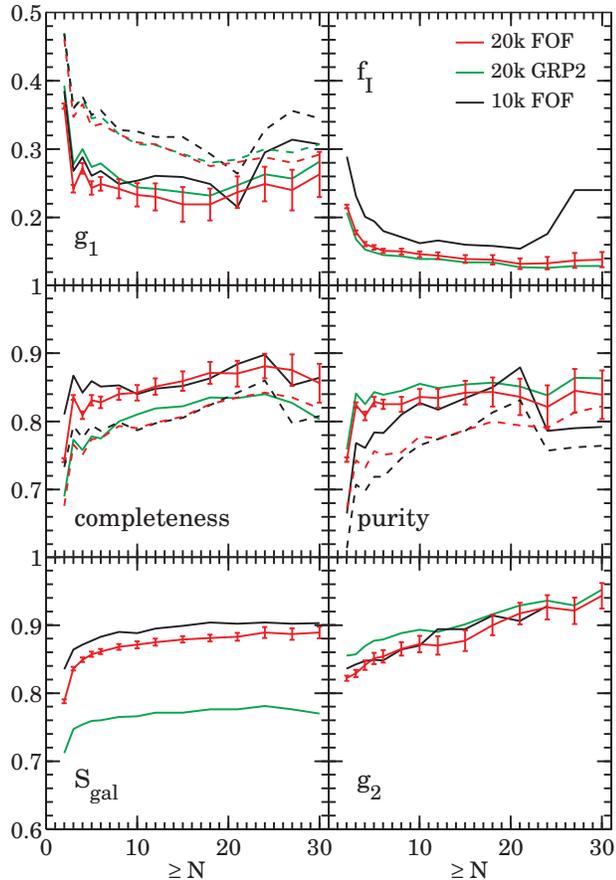}
	\caption{Cumulative statistics of the mock group catalogs as a function of observed richness $N$. The mean for the 20k FOF group catalogs is shown by the red lines, for the 20k GRP$_2$ group catalogs by the green lines, and for the FOF 10k group catalogs by the black lines. \emph{Upper left panel}: the solid lines correspond to $g_1$ and the dashed lines to $\tilde g_1$. \emph{Upper right panel:} interloper fraction $f_{\rm I}$. \emph{Middle panels:} the solid lines correspond to $c_1$ (left) and $p_1$ (right) and the dashed lines to $c_2$ and $p_2$, respectively. \emph{Lower left panel:} galaxy success rate $S_{\rm gal}$. \emph{Lower right panel:} goodness $g_2$. In all panels, the error bars refer to the standard deviation of the mean. For the sake of clarity they are only shown for the 20k FOF catalogs.\label{fig2:catalog_statistics}}
\end{figure}

\begin{figure}
	\centering
	\includegraphics[width=0.45\textwidth]{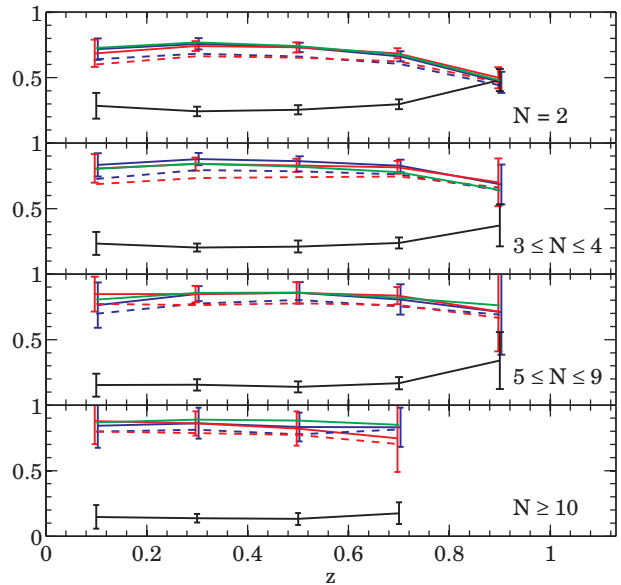}
	\caption{Statistics for the FOF mock group catalogs as function of redshift. The four panels show different richness classes as indicated by the labels. The solid curves indicate the completeness $c_1$ (blue), purity $p_1$ (red), galaxy success rate $S_{\rm gal}$ (green), and the interloper fraction $f_{\rm I}$ (black). The dashed lines correspond to $c_2$ (blue) and $p_2$ (red). The error bars are only shown for $c_1$, $p_2$ and $f_{\rm I}$ for clarity and correspond to the standard deviation among the 24 mock catalogs. The robustness of the catalog statistics over most of the redshift range is clear.\label{fig:statistics_z}}
\end{figure}

\begin{figure}
	\centering
	\includegraphics[width=0.45\textwidth]{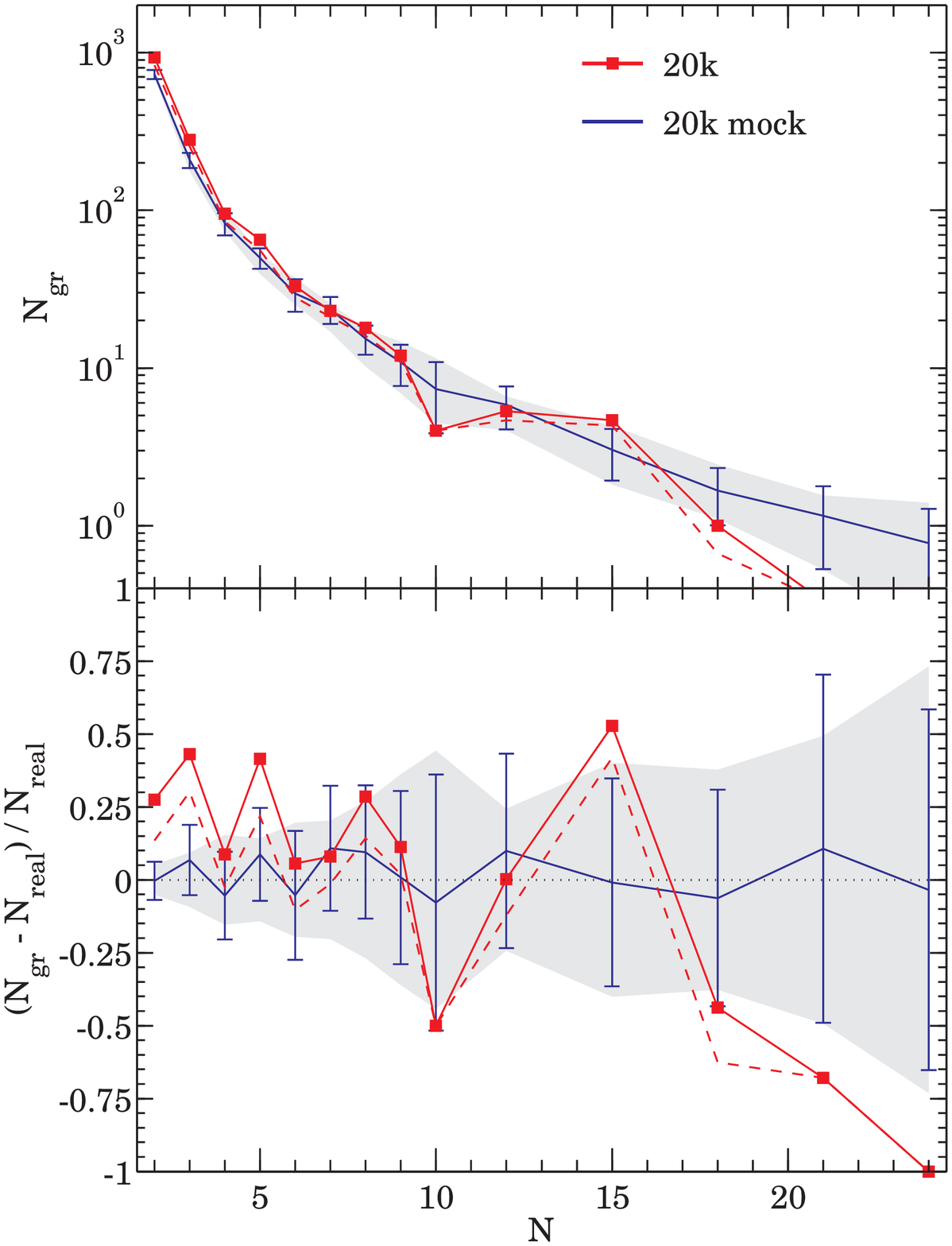}
	\caption{Number of groups $N_{\rm gr}$ as a function of observed richness $N$. The upper panel shows the absolute number of groups and the lower panel the relative number compared to the number of real groups within the mock catalogs. Shown are the 20k FOF catalog (red solid line), the 20k GRP$_2$ catalog (red dashed line), and the mean of the 20k FOF mock catalogs (blue line). The error bars indicate the standard deviation among the 24 mock catalogs, and the gray shaded area corresponds to the sample variance for the real 20k mock groups. \label{fig2:n_of_N}}
\end{figure}

The global properties of the 20k mock group catalogs are summarized in Table \ref{tab2:catalog_properties} and in Figures \ref{fig2:catalog_statistics}-\ref{fig2:different_associations}. If the pairs are excluded, the full catalogs exhibit a completeness $c_1 \gtrsim 83 \%$ and a purity $p_1 \gtrsim 83 \%$ for any richness. If we restrict the sample to the central region and to the redshift range $0.1 < z < 0.8$, where most groups are, the completeness for these groups even rises to $c_1 \gtrsim 85 \%$, while the purity remains about the same as before.
\begin{deluxetable}{ccccc}
%\tablewidth{0.7\textwidth}
\tablewidth{0pt}
\setlength{\tabcolsep}{4mm}
\tablecaption{Statistics of the 20k mock group catalogs for different observed richness ranges $N$}
\tablehead{
	\colhead{}&
  \colhead{$N = 2$} &
  \colhead{$3 \leq N \leq 4$} &
  \colhead{$5 \leq N \leq 9$} &
  \colhead{$N \geq 10$}
  }
\startdata

\sidehead{Full field and full redshift range}

$c_1$          & $0.69 \pm 0.02$ & $0.84 \pm 0.03$ & $0.83 \pm 0.04$ & $0.84 \pm 0.06$ \\
$c_2$          & $0.62 \pm 0.02$ & $0.76 \pm 0.03$ & $0.77 \pm 0.04$ & $0.80 \pm 0.08$ \\   
$p_1$          & $0.69 \pm 0.02$ & $0.82 \pm 0.02$ & $0.83 \pm 0.04$ & $0.84 \pm 0.06$ \\
$p_2$          & $0.63 \pm 0.02$ & $0.74 \pm 0.03$ & $0.75 \pm 0.04$ & $0.78 \pm 0.06$ \\
$S_{\rm gal}$  & $0.70 \pm 0.02$ & $0.80 \pm 0.02$ & $0.84 \pm 0.02$ & $0.87 \pm 0.02$ \\
$f_{\rm I}$          & $0.30 \pm 0.02$ & $0.22 \pm 0.02$ & $0.17 \pm 0.02$ & $0.15 \pm 0.02$\\\\

\hline
\sidehead{Central region and $0.1 < z < 0.8$}

$c_1$          & $0.72 \pm 0.02$ & $0.85 \pm 0.03$ & $0.84 \pm 0.05$ & $0.86 \pm 0.06$ \\
$c_2$          & $0.65 \pm 0.02$ & $0.78 \pm 0.04$ & $0.78 \pm 0.05$ & $0.81 \pm 0.06$ \\   
$p_1$          & $0.72 \pm 0.02$ & $0.82 \pm 0.03$ & $0.85 \pm 0.04$ & $0.83 \pm 0.05$ \\
$p_2$          & $0.64 \pm 0.02$ & $0.73 \pm 0.03$ & $0.77 \pm 0.04$ & $0.78 \pm 0.07$ \\
$S_{\rm gal}$  & $0.73 \pm 0.02$ & $0.81 \pm 0.03$ & $0.85 \pm 0.03$ & $0.88 \pm 0.02$ \\
$f_{\rm I}$          & $0.27 \pm 0.02$ & $0.22 \pm 0.02$ & $0.16 \pm 0.02$ & $0.14 \pm 0.02$

\enddata
\tablecomments{The numbers refer to the mean and the error bars to the standard deviation among the 24 mock catalogs.}

\label{tab2:catalog_properties}
\end{deluxetable}

In Figure \ref{fig2:catalog_statistics} the cumulative statistics of the 20k FOF mock catalogs are shown (red line) and compared those of the 10k mock catalogs (black line) and the 20k GRP$_2$ mock subcatalogs (green line), i.e., all groups with a GRP$_2 > 0$.
From the $g_1$-panel it is clear that compared to the 10k catalogs the 20k catalogs constitute an improvement of about 5\%-10\% which is significant in terms of the statistical error of the mean of the 24 mock catalogs which presumably reflects the range of things (such as overapping groups, spatial distribution of galaxies in groups etc.) which can influence the purity and completeness. This superiority is less obvious from a glance at the completeness and the purity (middle panels). For $N \gtrsim 10$ the completeness of the catalogs, both $c_1$ and $c_2$, are similar while the purity of the 20k catalogs is slightly higher and the overall line is much more uniform over a broad range for $N$. For $N \lesssim 10$ the completeness of the 10k catalogs is higher, but this deficiency is more than balanced by the improved purity of the 20k catalogs. The trends of the galaxy success rate $S_{\rm gal}$ and the interloper fraction $f_{\rm I}$ are similar between the 10k and 20k. The 20k catalogs have significantly less interlopers for all $N$.

Overall, the 20k mock group catalogs are generally purer than the 10k mock catalogs. In fact, they are so pure that, as already noted above, there is almost no difference between the FOF and the corresponding GRP$_1$ subcatalogs.  As expected, the GRP$_2$ catalogs are even purer than the FOF ones, but at the expense of completeness. While the $g_1$ goodness of the GRP$_2$ catalog is worse than that of the FOF catalogs, the $g_2$ goodness is better for groups $N \lesssim 10$. Thus, selecting only the GRP$_2$ groups slightly diminishes the contamination from overmerging and fragmentation.

The catalog statistics as a function of redshift are shown in Figure \ref{fig:statistics_z} for different richness classes.
It is clear that all statistics are fairly robust over the whole redshift range for any richness of group, as was already demonstrated for the 10k sample (cf.~Fig.~9 of K09). Only at the very high redshift end $z > 0.8$ and for the smallest groups is a weak redshift dependence apparent.

The superiority of the 20k catalogs over the 10k catalogs can, however, only be partially assessed by Figure \ref{fig2:catalog_statistics}. One of its major successes is that the new catalogs correctly reproduce the number of  groups as a function of richness $N$. Figure \ref{fig2:n_of_N} shows the relative abundance of reconstructed groups $N_{\rm rec}$ (lower panel blue line) compared to real groups $N_{\rm real}$ in the mock catalogs.
It is clearly seen that the mean number of reconstructed groups follow extremely well the number of real groups for all $N$. Even the scatter in reconstructed groups among the 24 mock catalogs is well within the sample variance of the real groups. Note that in K09, it was the 1WM subcatalogs that had this property, while the basic FOF multi-run catalogs contained rather too many groups for small $N$ (see Fig.~6 of K09).

\section{The spectroscopic group catalog}\label{sec:spectroscopic_group_catalog}

The group catalog produced with the actual zCOSMOS 20k sample is given in Tables \ref{tab:groups} and \ref{tab:galaxies}. The first table provides a list of all groups along with their properties and the second the corresponding group galaxy sample containing the spectroscopic and photometric group population. Regarding the construction of the photometric group population we refer to Section~\ref{ch:photometric_group_population_method}. In the following, we will call the actual zCOSMOS FOF group catalog just the ``20k group catalog''. The positions of the 20k groups in redshift space are shown in Figure \ref{fig2:cone}.
%\textbf{The light cone of the 20k groups is shown in Figure \ref{fig2:cone}}. 
\begin{deluxetable*}{cccccccrcc}
%\tablewidth{0.6\textwidth} 
%\tablecolumns{10}
\tablewidth{0pt}
\tablecaption{The zCOSMOS 20k group catalog (excerpt)}
\tablehead{
  \colhead{Group-ID} &
  \colhead{$N$\tablenotemark{a}} &
   \colhead{$N_{\rm corr}$\tablenotemark{b}} &
  \colhead{$\alpha_{\rm gr}$\tablenotemark{c}} &
  \colhead{$\delta_{\rm gr}$\tablenotemark{c}} &
  \colhead{$z_{\rm gr}$\tablenotemark{d}} &
  \colhead{$r_{\rm fudge}$\tablenotemark{e}} &  
  \colhead{$\hat \sigma$\tablenotemark{f}} &
  %\colhead{$\log(M_{\rm fudge}/M_\odot)$\tablenotemark{c}} &
   \colhead{$\log \left(\frac{M_{\rm fudge}}{M_\odot}\right)$\tablenotemark{g}} &
  \colhead{GRP$_2$\tablenotemark{h}} \\
  \colhead{} &
    \colhead{} &
    \colhead{} &
  \colhead{(deg)} &
  \colhead{(deg)} &
  \colhead{} &
  \colhead{(Mpc)} &
  \colhead{(km s$^{-1}$)} &
  \colhead{} &
  \colhead{} 
}
\startdata
       0  &  14   & 33&   150.02209  &    2.01328 &      0.0787  &   0.646 &  433  &  13.51  &  0.93  \\
       1  &  30   & 54&  150.35758  &   2.44265  &     0.1230   &  0.652  & 454     &13.56   & 0.63   \\
       2  &  33   & 52&  149.86613   &   1.76547  &     0.1245   &  0.674  & 587     &13.52   & 0.61  \\
       3  &  14   & 28&  150.42153  &   2.44418  &     0.2160   &  0.532  & 298    &13.45  &  1.00  \\
       4  &  14   & 97&  150.20008   & 1.65232    &   0.2202    & 0.722  &1008     &13.69   & 0.93  \\
       5  &  17    & 36&  150.10545  &  2.36170   &    0.2201   &  0.577  & 745   &13.44   & 0.94  \\
       6  &  20    & 27&  150.45635   &  2.68079   &    0.2186    & 0.515  & 642   &13.42   & 0.95   \\
       7  &  17    & 28&  150.04641   &  2.43245   &    0.2200    & 0.532  & 662   &13.40   & 0.71    \\
       8  &  15    & 16&  150.23142   &   2.55729  &     0.2199  &   0.627  & 418     &13.49  &  0.87 
\enddata
\tablecomments{This table is available in its entirety in a machine-readable form in the online journal. A portion is shown here for guidance regarding its form and content.}
\tablenotetext{a}{Number of spectroscopic members.}
\tablenotetext{b}{Corrected richness with respect to the flux limit (see Sect.~\ref{sec:corrected_richness}).}
\tablenotetext{c}{Improved group centers defined in Section~\ref{sec:group_centers}.}
\tablenotetext{d}{Mean redshift of the spec-$z$ group members.}
\tablenotetext{e}{Fudge radius in physical Mpc (see Sect.~\ref{sec:physical_properties}).}
\tablenotetext{f}{Velocity dispersion for groups with $N \geq 5$ (see Sect.~\ref{sec:physical_properties}).}
\tablenotetext{g}{Fudge mass for the DM halo (see Sect.~\ref{sec:physical_properties}).}
\tablenotetext{h}{Group purity parameter (GPR$_2$) (see Sect.~\ref{sec:sub_catalogs}).}

\label{tab:groups}
\end{deluxetable*}
\begin{deluxetable*}{ccccccccccc}
\tablecaption{Spec-$z$ and photo-$z$ group galaxies (excerpt)}
\tablehead{
  \colhead{Galaxy ID} &
  \colhead{Group ID} &
  \colhead{20k Flag\tablenotemark{a}} &
    \colhead{GAP$_2$\tablenotemark{b}} &
  \colhead{$\alpha$} &
  \colhead{$\delta$} &
  \colhead{$z$\tablenotemark{c}} &
    \colhead{$\log(M_\ast/M_\odot)$\tablenotemark{d}} &
  \colhead{$p$\tablenotemark{e}} &
  \colhead{$p_{\rm M}$\tablenotemark{f}} &
    \colhead{$p_{\rm MA}$\tablenotemark{g}}
    \\
  \colhead{} &
  \colhead{} &
    \colhead{} &
      \colhead{} &
  \colhead{(deg)} &
  \colhead{(deg)} &
     \colhead{} &
   \colhead{} &
    \colhead{} &
        \colhead{} &
  \colhead{}
}
\startdata
    819041  &   0 &     1 &  1   &   149.99837  &  2.03514  &    0.0789 &     9.33  &  0.92  &  0.00   &     0.00 \\    
    818934  &   0 &     1 &  1   &   150.02406  &  1.96865  &    0.0779 &    8.30   &  0.89  &  0.00   &     0.00   \\
    818888  &   0 &     1 &  1   &   150.03653  &  2.02487  &    0.0794 &    7.85   &  0.96  &  0.00   &     0.00    \\
    819026  &   0 &     1 &  1   &   150.00038  &  1.97859  &    0.0802 &   10.05  &  0.90  &  0.00   &     0.00   \\
    818839  &   0 &     1 &  1   &   150.04871  &  2.07792  &    0.0775 &     8.17  &  0.82  &  0.00   &     0.00    \\
    819133  &   0 &     1 &  1   &   149.96812  &  2.06726  &    0.0779 &     8.17  &  0.80  &  0.00   &     0.00   \\
    819032  &   0 &     1 &  0   &   149.99948  &  1.98699  &    0.0805 &   10.22  &  0.92  &  0.00   &     0.00    \\
    819060  &   0 &     1 &  1   &   149.99123  &  1.99116  &    0.0797 &     8.11  &  0.91  &  0.00   &     0.00    \\
    818935  &   0 &     1 &  1   &   150.02393  &  2.07273  &    0.0779 &     7.97  &  0.85  &  0.00   &     0.00    \\
    818815  &   0 &     1 &  1   &   150.05394  &  2.03343  &    0.0785 &     8.47  &  0.91  &  0.00   &     0.00    \\
    819118  &   0 &     1 &  1   &   149.97241  &  2.10540  &    0.0781 &     7.99  &  0.71  &  0.00   &     0.00   \\
    819104  &   0 &     1 &  1   &   149.97723  &  2.00483  &    0.0779 &   10.16  &  0.89  &  0.00   &     0.00    \\
    818982  &   0 &     1 &  1   &   150.01341  &  2.02956  &    0.0791 &   10.70  &  0.96  &  0.16   &     0.96    \\
    818787  &   0 &     1 &  1   &   150.06047  &  2.00672  &    0.0785 &   10.48  &  0.91  &  0.01   &     0.00    \\
    700213  &   0 &     0 & $-1$  &   150.07021  &  1.85821  &    0.1029 &     8.19  &  0.19  &  0.00   &     0.00   \\ 
    700241  &   0 &     0 & $-1$  &   149.98257  &  1.80462  &    0.0964 &     7.87  &  0.09  &  0.00   &     0.00    
\enddata

\tablecomments{This table is available in its entirety in a machine-readable form in the online journal. A portion is shown here for guidance regarding its form and content.}
\tablenotetext{a}{1 if spec-$z$ is available, otherwise 0.}
\tablenotetext{b}{Galaxy purity parameter for spec-$z$ members (see Sect.~\ref{sec:sub_catalogs}), $-1$ for photo-$z$ members.}
\tablenotetext{c}{Spec-$z$ if available, otherwise photo-$z$.}
\tablenotetext{d}{Stellar mass (computed without considering mass return, see Sect.~\ref{sec:photo}).}
\tablenotetext{e}{Association probability (see Sect.~\ref{sec:p}).}
\tablenotetext{f}{Probability to be the most massive (see Sect.~\ref{sec:p_M}).}
\tablenotetext{g}{See Sect.~\ref{sec:central_galaxy}.}

\label{tab:galaxies}
\end{deluxetable*}

The basic properties of the 20k group catalog are summarized in Table~\ref{tab:GRP} and compared to the 10k catalog. For $N \geq 2$, the 20k catalog contains 1496 groups, almost twice as many as the 10k catalog, while it has four times as many groups with $N \geq 10$. 
\begin{deluxetable*}{crcccrcc}
\tablewidth{0pt}
\tablecaption{Basic statistics for the zCOSMOS 10k and 20k group catalogs}
\tablehead{
\colhead{} & \multicolumn{3}{c}{10k} & \colhead{} & \multicolumn{3}{c}{20k} \\ 
\cline{2-4} \cline{6-8} \\ 
  \colhead{} &
  \colhead{$N_{\rm gr}$\tablenotemark{a}} &
  \colhead{$f_{\rm GRP_1}$\tablenotemark{b}} &
  \colhead{$\langle\rm{GRP}_1\rangle$} &&
  \colhead{$N_{\rm gr}$\tablenotemark{a}} &
  \colhead{$f_{\rm GRP_2}$\tablenotemark{b}} &
  \colhead{$\langle\rm{GRP}_2\rangle$}
}

\startdata
$N = 2$           & 514 & 0.79 & 0.79 && 932 & 0.89 & 0.89\\
$3 \leq N \leq 4$ & 184 & 0.81 & 0.77 && 374 & 0.91 & 0.89\\
$5 \leq N \leq 9$ & 91  & 0.95 & 0.87 && 151 & 0.87 & 0.81\\
$N \geq 10$       & 11  & 1.0  & 0.93 &&  41 & 0.90  & 0.78
\enddata

\tablenotetext{a}{Number of groups.}
\tablenotetext{b}{Fraction of groups in the corresponding GRP$_i$, $i \in \{1,2\}$, subcatalog.}

\label{tab:GRP}
\end{deluxetable*}

\subsection{Group robustness}\label{sec:group_robustness}

One of the main prerequisites for estimating the properties of reconstructed groups is the fact that the group is reliably identified. If the group is overmerged or fragmented, the derived properties such as mass or physical size will be severely affected and may have little or nothing to do with those of the real group. For reconstructed groups that do \emph{not} have a 2WM to their real groups, we cannot even, in general, perform a unique one-to-one comparison between the properties of real groups and those of the reconstructed groups. This again emphasizes the importance for a group catalog to be not only optimal in respect to the one-way statistics $c_1$ and $p_1$, but also regarding the two-way statistics $c_2$ and $p_2$. 

\begin{figure}
	\centering
	\includegraphics[width=0.46\textwidth]{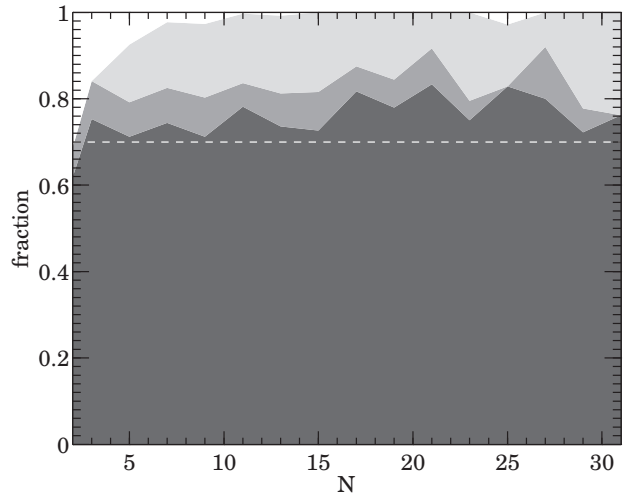}
	\caption{Portions of the different possible associations between reconstructed and real groups for the 20k mock catalogs as a function of observed richness $N$. The bottom layer shows the fraction of reconstructed groups having 2WM to real groups. The surface of this layer is equal to $p_2(N)$. The Dark gray layer shows the portion of 1WM (which are no 2WM) from reconstructed groups to real groups (``fragmentation'') and the light gray layer the corresponding portion from real groups to reconstructed groups (``overmerging''). The white area corresponds to the portion for which no association exists (``spurious groups''). The dashed line is a benchmark at 0.7. \label{fig2:different_associations}}
\end{figure}

\begin{figure*}
	\centering
	\includegraphics[width=0.8\textwidth]{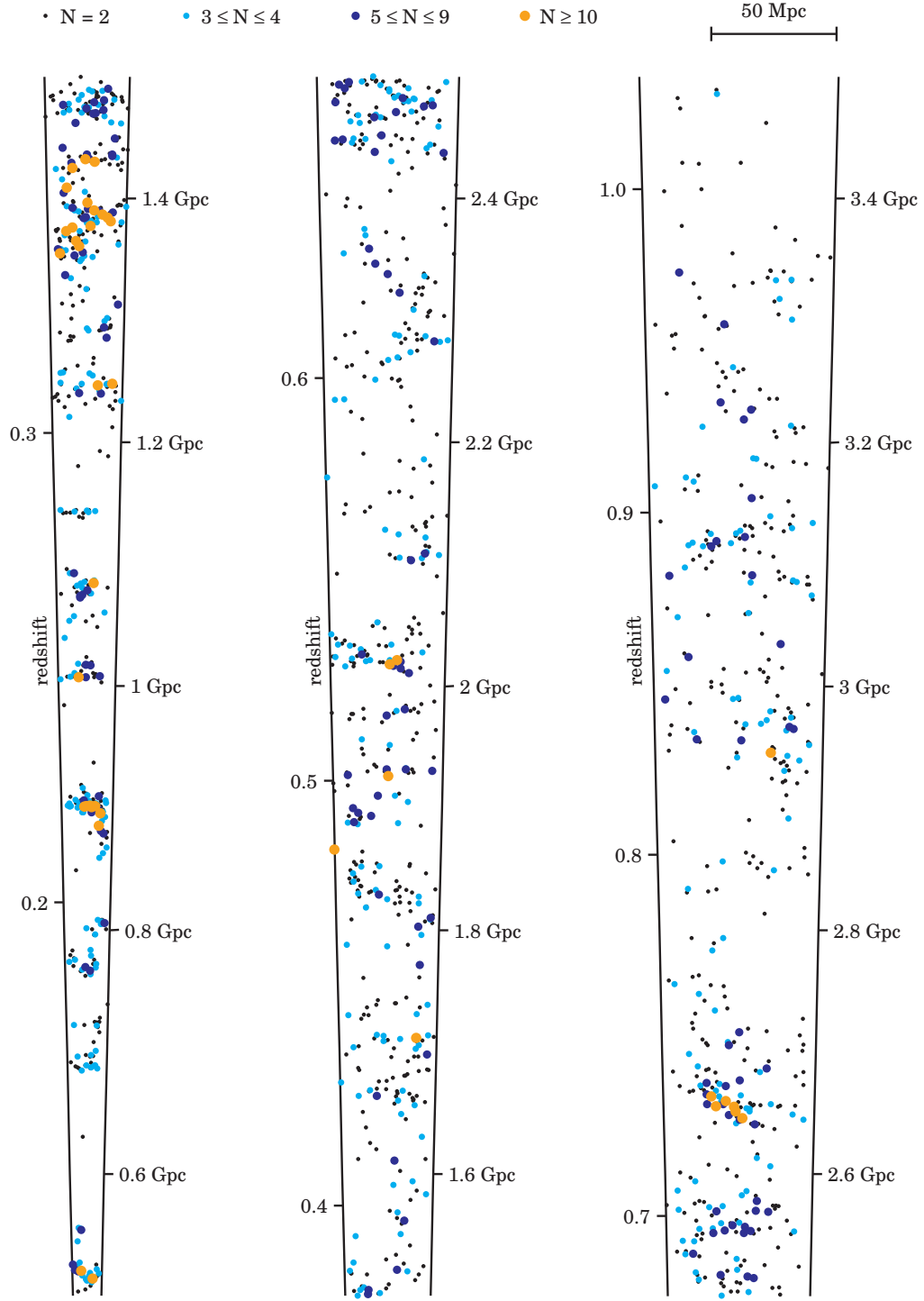}
	\caption{Positions of the zCOSMOS 20k groups in redshift space. The groups are plotted as a function of right ascension and comoving distance, where the richness $N$ of the groups is color coded as indicated above the cone. The labels on the left side of the cone indicate the redshift and the ones on the right side the corresponding comoving distance.  Note that the transverse scale of the cone has been stretched by about a factor of two for clarity. In reality, the comoving depth of this cone (from $z = 0.1$ to 1) is about 70 times longer than its transverse comoving size at $z = 0.5$. The comoving transverse scale of the cone is indicated by the horizontal bar at the top. The clustering of the groups and the cosmic large-scale structure are clearly visible up to the highest redshifts.\label{fig2:cone}}
\end{figure*}

Figure \ref{fig2:different_associations} shows the fractions of groups as a function of observed richness $N$ that have the four different kinds of possible associations: full 2WM, a 1WM from reconstructed to real (i.e., fragmentation), a 1WM in the opposite direction (i.e., overmerging), and no association at all.
The percentage of reconstructed groups exhibiting a 2WM to real groups is $\gtrsim\! 75 \%$. Of the remaining $\sim \!25\%$, the fraction of overmerged groups is higher than that of fragmented groups. It should also be noted that for groups with $N \gtrsim 5$ there are almost no spurious groups. That is, essentially every group that is found constitutes a real physical structure in the universe, but in 20\%-30\% of cases, the group-finder has made it significantly too small or too big (by a factor of more than two in membership) compared to the real group. The fact that Figure \ref{fig2:different_associations} is basically independent of $N$ is a consequence of the application of the multi-run scheme (see Sect.~\ref{sec:optimization_strategy}).

Since the FOF groups depend solely on the two quantities $l_{\rm per}$ and $l_{\rm par}$ which are the linking-lengths perpendicular and parallel to the line of sight, respectively, a natural question is whether a given group is sensitive to the particular choice of these linking lengths, or whether slightly different values would not significantly alter the resulting group? To answer this question we have introduced a ``group robustness'' parameter, $f_{\rm rob}(f)$ for each group, by running the groupfinder with the linking-lengths $f \cdot l_{\rm per}$ and $f \cdot l_{\rm par}$, parameterized by the scale factor $f$, and computing for each group the fraction 
\begin{equation}
f_{\rm rob}(f) = \left\{\begin{array}{ll} N/N(f)\:,& \quad\text{if } f < 1\\N(f)/N\:,& \quad\text{if } f \geq 1 \:,\end{array}\right.
\end{equation}
where $N(f)$ is the new richness of that group. This assures that $f_{\rm rob}(f)$ takes only values between 0 and 1 and that the robustness increases for higher $f_{\rm rob}(f)$, with $f_{\rm rob}(f) = 1$ being a highly robust structure. $f_{\rm rob}(f)$ is a measure of how sensitive the implied membership is to changes in the linking length. For $f < 1$ it probes the robustness in respect to fragmentation and for $f > 1$ in respect to overmerging.  

\begin{figure}
	\centering
	\includegraphics[width=0.46\textwidth]{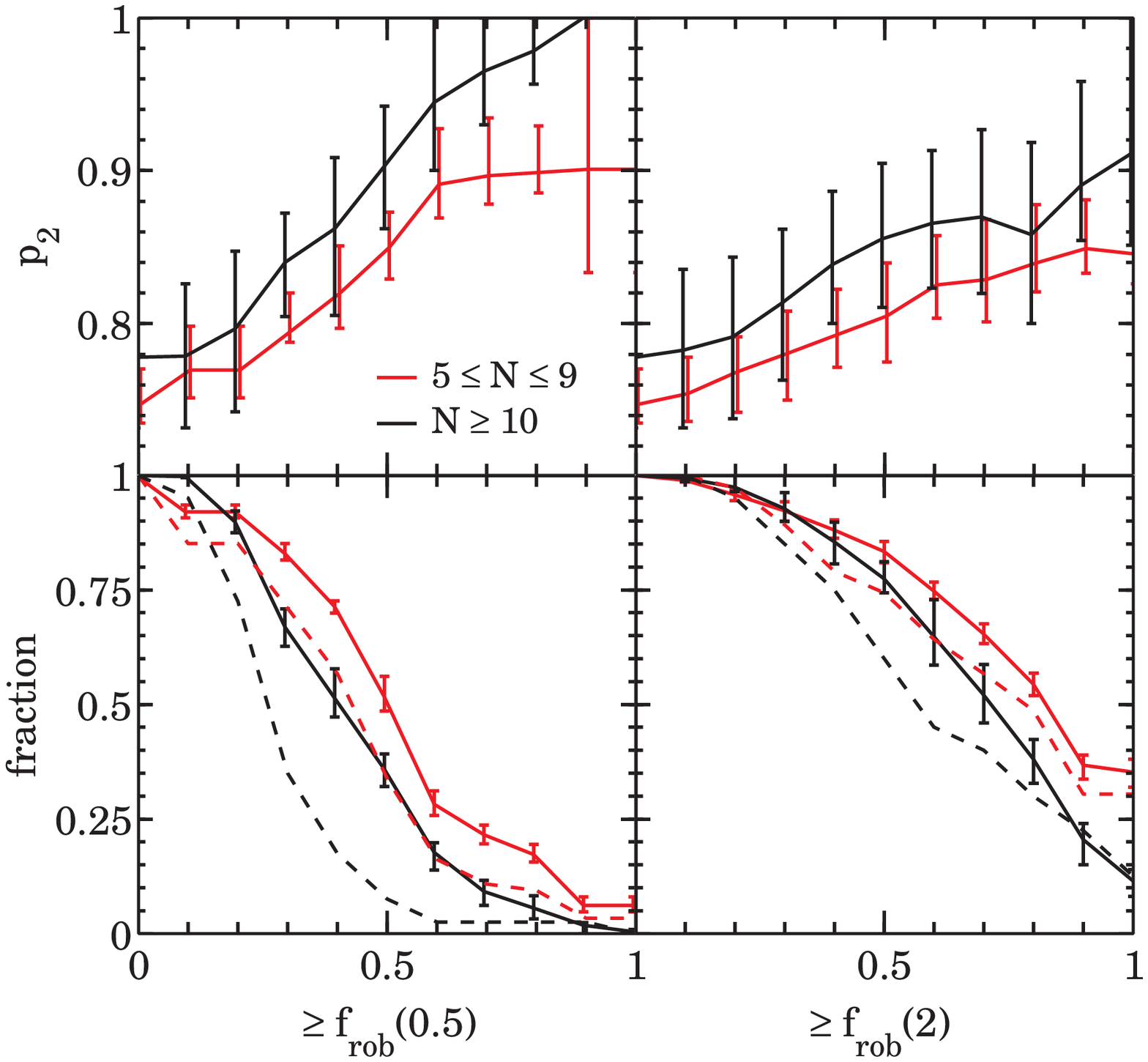}
	\caption{Purity and group fraction as a function of group robustness. The solid lines show the median and the error bars the upper and lower quartiles of the 24 20k mock catalogs. The dashed lines are the results for the actual 20k group catalog. Red lines correspond to the richness class $5 \leq N \leq 9$ and the black lines to $N \geq 10$.\label{fig2:group_robustness}}
\end{figure}

Figure \ref{fig2:group_robustness} shows the results for $f = 0.5$ and $f = 2$ for 20k FOF mock groups in the richness classes $5 \leq N \leq 9$ (red lines) and $N \geq 10$ (black line). These results are not sensitive to the precise value of $f$.  
The upper panels exhibit the $p_2$ statistics for the corresponding $f_{\rm rob}(f)$ selected subsamples.   Reducing the linking length tends to have a bigger effect than increasing it. Roughly 50\% of groups in the mock catalogs lose a half of their members when the linking lengths are halved, but only 25\% of groups double their memberships when the linking lengths are doubled. As would be expected, the overall purity increases strongly as the robustness $f_{\rm rob}(f)$ approaches unity, for both $f$ being larger and smaller than one, and groups whose membership is stable to changes in the linking lengths are, not surprisingly, likely to be the purest.  However, the lower panels make clear that raising the purity of subsamples significantly by applying cuts in the robustness comes at the expense of losing many groups.

For the actual 20k group catalog the group fraction is shown by the dashed lines in Figure \ref{fig2:group_robustness}. Particularly the big groups are significantly less robust in respect to fragmentation than the corresponding mock groups. We do not know the reason for this but it matches other properties of the 20k group catalog such as the lack of high richness groups (see Sect.~\ref{seq:n_of_N}). Note that in contrast to the completeness and purity, the group robustness is one of the few quantities that can be computed using the actual data without the need for mock catalogs and thus allows a direct comparison with simulated data.

\subsection{Estimates of physical properties}\label{sec:physical_properties}

As pointed out in K09, we are able to estimate the velocity dispersion $\sigma_{\rm v}$ for groups with $N \geq 5$ and $\sigma_{\rm v} \gtrsim 350$ km s$^{-1}$ to an accuracy of about 25\%. On the other hand, a reliable estimation of dynamical mass by means of the virial theorem has proved to be very difficult, not only because of the error of the velocity dispersion enters the virial theorem quadratically, but also because reliable estimates of the virial radius are very hard to obtain. Using the mock catalogs we found that the projected apparent extension of a group hardly correlates at all with the virial radius of the corresponding DM halo. The unavailability of reliable dynamical mass estimates is one major shortcoming of our group catalog, and others constructed in similar ways. To have at least an idea of the typical mass of the groups we introduced in K09 the so-called fudge mass by taking the corrected richness $\tilde N$ of the group (i.e., observed richness $N$ corrected for SSR and RSR) at a given redshift $z$ as a proxy for its mass.

In the same spirit we can define ``fudge quantities'' $Q_{\rm fudge}$ for any quantity $Q$ that at a given redshift exhibits a correlation to the corrected richness $\tilde N$ or to another quantity $\tilde Q$ which is independently measurable (e.g., velocity dispersion, projected extension). That is, a group at redshift $z$ with corrected richness $\tilde N$ and with the measured property $\tilde Q$ can be assigned a corresponding $Q_{\rm fudge}$ defined by
\begin{equation}
 Q_{\rm fudge} = \left\langle Q_{\rm mock}(\tilde N,z,\tilde Q) \right\rangle\:,
\end{equation}
where the brackets $\langle\rangle$ denote the average considering all reconstructed groups with 2WM to real groups for which the corresponding measured quantities are within some range of $\tilde N$,$z$ and $\tilde Q$, and $Q_{\rm mock}$ denotes the correct group property of the corresponding real group.

Additionally to the fudge mass we have computed fudge estimates for the halo virial velocity (``fudge velocity'') and halo radius (``fudge radius''). For the fudge velocity we have used the apparent velocity dispersion $\sigma_{\rm v}$ as $\tilde Q$ and for the fudge radius, we use the apparent projected size of the group, as defined below. The scatter of the estimated quantities compared to the true quantities for reconstructed 20k mock groups exhibiting a 2WM to real groups is given in Table \ref{tab2:galaxy_properties}. As expected the errors decrease with increasing observed richness $N$. Note that the fudge quantities must not be mistaken for real physical estimates of the corresponding quantity, they are rather ``typical'' values calibrated using the mock catalogs.
\begin{deluxetable}{llrcccc}
%\tablewidth{0.7\textwidth}
\tablewidth{0pt}
\tablecaption{Errors for the fudge quantities using the 20k mock catalogs}
\tablehead{
	\colhead{Quantity} &
  \colhead{Error} &
  \colhead{$N = 2$} &
  \colhead{$3 \leq N \leq 4$} &
  \colhead{$5 \leq N \leq 9$} &
  \colhead{$N \geq 10$} 
  }
\startdata

$M_{\rm fudge}$ & $\Delta$ dex & 0.37 & 0.27 & 0.18 & 0.15 \\
$v_{\rm fudge}$ & Rel. error & 21\% & 19\% & 13\% & 9\% \\
$r_{\rm fudge}$ & Rel. error & 27\% & 23\% & 16\% & 11\%

\enddata

\label{tab2:galaxy_properties}
\end{deluxetable}

\subsection{Number of groups as a function of $N$} \label{seq:n_of_N}

The most straightforward way to compare the actual data with the mock data is by means of the number of reconstructed groups as a function of observed richness $N$. This is shown in Figure~\ref{fig2:n_of_N}. Compared to the mock data the number of groups of the 20k group catalog is mostly within the range expected due to sample variance within the 24 mock catalogs. Since for the 20k mock catalogs the number of reconstructed groups traces very well the number of real groups for any richness, there is no need to distinguish between them.

The overall slope of the $N_{\rm gr}(N)$ function for the actual data, however, is steeper than for the mock data. Particularly the number of groups with two and three members is about 25\%-50\% higher than in the mock catalogs and for $N \gtrsim 18$ there is a significant lack of groups in the 20k sample compared with the mock catalogs. Both trends were already noted for the 10k sample in K09 and are now confirmed with the larger 20k sample. The excess of groups with $N \lesssim 3$ cannot be blamed to the existence of secondary objects (serendipitous observations in the spectroscopic slits) which could boost the number of small groups since the fraction of such objects is only about 2\%. Interestingly, a significant lack of high richness groups relative to the Millennium simulation has recently also been reported for the large GAMA FOF group catalog at local redshift \citep{robotham2011}, which indicates that this lack is not a  peculiarity of the COSMOS field.

It should be particularly noted that many of the individual mock catalogs contain groups which are much larger than those in zCOSMOS. While the largest group in the 20k sample has 33 members, there are on average about 3-4 groups with $N \geq 40$ per 20k mock catalog and 1-2 groups with $N \geq 60$. These huge groups are not an artifact of our group-finding algorithm, but are present as real groups in the mock catalog.  Since the high-mass end of the halo mass function is very sensitive to the amplitude of the matter power spectrum in the universe, $\sigma_8$, the large number of big groups in the mock catalogs could reflect the fact the $\sigma_8 = 0.9$ for the Millennium simulation is too large relative to recent measurements of $\sigma_8 \simeq 0.82\pm0.02$ \citep[e.g.,][]{komatsu2011}.

A direct measurement of $\sigma_8$ by means of the group mass function is, however, very difficult, for two main reasons. First, it is the high-mass end of the mass function that is most sensitive to $\sigma_8$ and where our catalog is most complete (see Fig.~\ref{fig2:halo_completeness}). Due to the relatively small volume of zCOSMOS we are in the regime of low number statistics for such high masses and thus are affected by cosmic variance, particularly at low redshift. Second, we checked that a mass cut by means of the fudge mass would introduce some mass-dependent systematics into the mass function estimation so that a robust estimation of $\sigma_8$ would require improved mass estimates. However, the fact that there is no group in the 20k group catalog with $M_{\rm fudge} > 2 \cdot 10^{14}\ M_{\odot}$, while there are $\sim\!3.5$ on average in each mock, certainly favors a low $\sigma_8$. Only three out of 24 mock catalogs (i.e., 12.5\%) contain no group with that high fudge mass.

At this point it is interesting to come back to the findings on the group robustness in Section~\ref{sec:group_robustness}. We noted that for big groups the group robustness in respect to fragmentation is significantly lower than for the corresponding mock groups (Fig.~\ref{fig2:group_robustness}, black dashed lines). This points in the same direction as the detected lack of big groups. There are not only fewer big groups in the zCOSMOS group catalog than in the mock catalogs, the observed groups are also less robust.

\subsection{Fraction of galaxies in groups}\label{sec:fraction_galaxies_groups}

A quantity closely related to the number of groups in a catalog is the fraction of galaxies that are in groups. Since the number of groups traces roughly the number of galaxies in zCOSMOS (cf.~Fig.~12 of K09), computing fractions of galaxies in groups instead of the absolute number of groups diminishes the effect of large-scale structure and associated cosmic variance. A measurement of this fraction allows further comparison with the mock catalogs and allows us to trace the buildup of the cosmic group environment over time. The analysis in this section will be entirely restricted to the central region of the zCOSMOS survey (see Fig.~\ref{fig:mask}).

\begin{figure}
	\centering
	\includegraphics[width=0.47\textwidth]{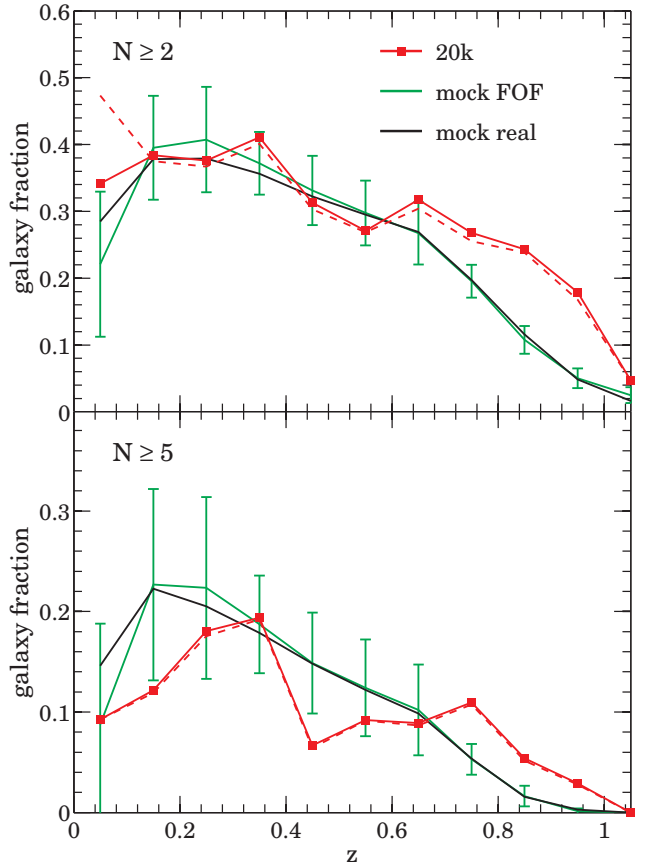}
	\caption{Fraction of galaxies in groups as a function of redshift for the whole flux-limited galaxy and group samples. The samples are restricted to the central region and show groups with $N \geq 2$ in the upper panel and $N \geq 5$ in the lower panel. The red solid line shows the fraction of galaxies in zCOSMOS 20k groups and the red dashed line the corresponding fraction if only groups are considered which are detectable without the existence of secondary objects. The black line shows the mean fraction of galaxies in real 20k mock groups and the green line the mean fraction of galaxies in reconstructed 20k mock groups. The error bars indicate the standard deviation among the 24 mock catalogs.  The mock catalogs are in fair agreement with the actual data for $z \lesssim 0.6$, but contain significantly too few groups for $z \gtrsim 0.6$.\label{fig:fraction_in_groups}}
\end{figure}

The fraction of galaxies in groups for the full flux-limited 20k group catalog and 20k galaxy sample is shown in Figure~\ref{fig:fraction_in_groups} as a function of redshift for $N \geq 2$ and $N \geq 5$.
The overall behavior of the fraction of galaxies in 20k groups (red line) matches quite well those of the reconstructed (or real) mock groups, at least in the redshift range $z \lesssim 0.6$.  At the highest redshifts, the fraction of group galaxies in zCOSMOS is significantly higher than in the mock catalogs. The reason for this is unclear. It may indicate a problem of the semi-analytic models to follow the evolution of galaxies. Most of these highest redshift groups are only detected as pairs, leading to possible worries about the sampling of objects. However, the red dashed line corresponds to groups which are still detectable even if all secondary objects were discarded, so this is not the cause of this effect. Furthermore, it should also be noted that the excess is also visible for much richer systems (lower panel in Fig.~\ref{fig:fraction_in_groups}).  It is noticeable (particularly for the lower panel) that the fraction of galaxies in groups is enhanced at the redshifts $z \sim 0.35$ and $z \sim 0.70$, where there are very large scale structures in the COSMOS field (cf.~Fig.~1 of K09 and Fig.~\ref{fig2:cone} in this paper). At low redshift the total fraction of galaxies in groups is about 40\%, which is consistent with the results from the low-redshift GAMA group catalog \citep{robotham2011}, despite the different limiting fluxes of the survey, presumably reflecting the weak dependence of satellite fraction on galaxy mass.

In order to get a clearer view into the buildup of the group environment over cosmic time, it is better to work with volume-limited samples of galaxies and groups, or as close approximations to such as can be constructed. We can approximate a volume-limited sample of galaxies by applying a cut in absolute magnitudes, chosen to evolve with redshift to deal, at least roughly, with the individual luminosity evolution of galaxies. We will apply the cut as
\begin{equation}\label{eq:m_b_z}
M_{\rm B} \leq M_{\rm B, lim} - z
\end{equation}
for different absolute magnitude limits $M_{\rm B, lim}$. We performed the analysis with three magnitude limits $M_{\rm B, lim}$ being $-19.75$, $-20.25$ and $-20.75$, respectively. The resulting galaxy populations are complete at least up to $z \sim 0.8$.

To construct a volume-limited sample of groups we select all groups with at least two members brighter than $M_{\rm B, lim} - z$. We use the observed richness rather than the richness corrected for SSR and RSR to avoid the scatter that is introduced by potentially large completeness corrections. This procedure is not perfect.  For instance, two galaxies may be linked at low redshift by others below the absolute magnitude cut, to form a ``group'' that would be undetected at high redshifts where the absolute magnitude limit is closer to the flux limit of the spectroscopic survey. This could lead to a redshift-dependent $c_1$ and/or $p_2$.  However, Figure \ref{fig:statistics_z} shows that the redshift dependence of $c_1$ and $p_1$ is negligible over the redshift range considered here.

\begin{figure}
	\centering
	\includegraphics[width=0.46\textwidth]{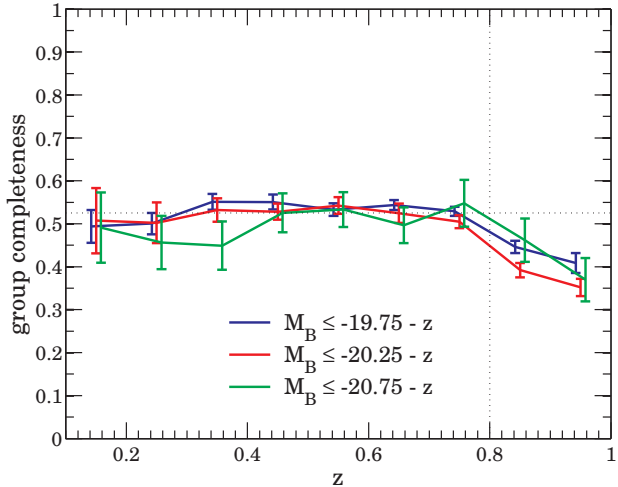}
	\caption{Completeness of mock groups containing at least two members brighter than $M_{\rm B, lim} - z$. The lines correspond to the mean of the 24 mock catalogs for different absolute magnitude limits (blue: $M_{\rm B, lim} = -19.75$; red: $M_{\rm B, lim} = -20.25$; green: $M_{\rm B, lim} = -20.75$) and the error bars to the standard deviation of the mean. It is obvious that for the redshift range $0.1 < z < 0.8$ the completeness for all three magnitude limits is fairly constant. \label{fig2:abs_mag_group_completeness}}
\end{figure}

To address these and other concerns, Figure \ref{fig2:abs_mag_group_completeness} shows the number of reconstructed mock groups compared with the number of all groups in the mock catalogs that host at least two bright galaxies, irrespective of whether these groups are detectable within the 20k mock samples, or not.  The obtained completeness is therefore lower than that shown in Figure \ref{fig2:catalog_statistics}, where only the ``detectable'' groups were considered as the parent sample.  The completeness computed in this way is found to be fairly constant in the redshift range $0.1 \lesssim z \lesssim 0.8$ for all three absolute magnitude cuts.
This reassures that there are no strong systematic biases for the absolute magnitude selected groups as a function of redshift.

\begin{figure}
	\centering
	\includegraphics[width=0.46\textwidth]{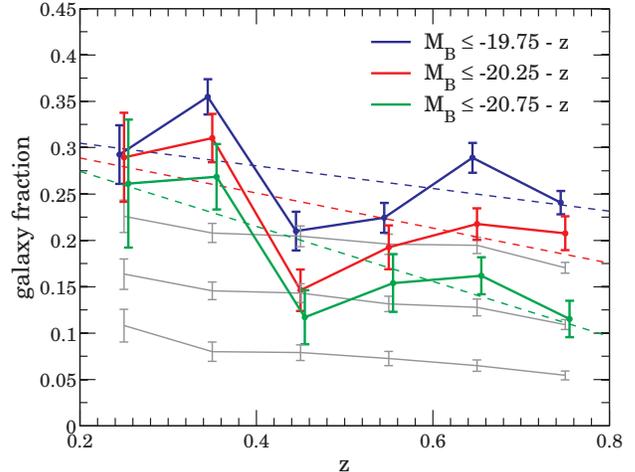}
	\caption{Fraction of galaxies in groups for volume-limited galaxy and group samples. The blue, red and green solid lines show the results for the zCOSMOS 20k data (blue: $M_{\rm B, lim} = -19.75$; red: $M_{\rm B, lim} = -20.25$; green: $M_{\rm B, lim} = -20.75$). The error bars for the actual data are obtained by bootstrapping and the dashed lines exhibit linear fits to the data points. The gray lines are the corresponding mean curves of the 24 FOF mock catalogs, where the luminosity increases for the lower curves. The error bars exhibit the standard deviation of the mean. This figure demonstrates the buildup of the cosmic group environment over the redshift range $0.2 < z < 0.8$. \label{fig2:group_fraction_abs_mag}}
\end{figure}

Having established that our ``volume-limited'' samples should be free of bias, the fraction of galaxies in the groups is shown in Figure \ref{fig2:group_fraction_abs_mag}.
For the 20k sample, we again see the signatures of the big structures at redshifts $z \sim 0.35$ and $z \sim 0.70$, as in Figure \ref{fig:fraction_in_groups}. This could indicate that the luminosity function of galaxies in groups is possibly environment dependent. Nevertheless, there is a clear overall trend for the fraction of (volume-limited) galaxies in (volume-limited) groups to significantly increase with decreasing redshift, as indicated by the dashed lines. This demonstrates the buildup of the cosmic group environment over a large fraction of the last 7 billion years. It should be noted that this result is insensitive of the precise form of the redshift correction in Equation~(\ref{eq:m_b_z}). 

Curiously, the observed fraction of (bright) galaxies in the 20k groups is significantly higher than in the mock groups. This finding is independent whether the flux limit for the mock catalogs is adjusted or not (see Sect.~\ref{sec:mocks}). The fraction in the mock catalogs, however, approaches that of the 20k sample as we go to fainter galaxies at the flux limit (in agreement with Figure \ref{fig:fraction_in_groups}). This suggests that the cause of the discrepancy on Figure \ref{fig2:group_fraction_abs_mag} could be a problem with the magnitudes of bright galaxies in the COSMOS mock light cones.

\section{Photometric group members}\label{ch:photometric_group_population_method}

For some applications it is very useful to have a complete galaxy sample down to a magnitude limit. For example, for studying the most massive galaxies in groups it must be ensured that these galaxies are present in the sample. Since even the 20k sample is only complete to about $55\%$, the spectroscopic group catalog is not yet optimized for this kind of studies. On the other hand, since zCOSMOS is performed on the COSMOS field which was followed in many wavelength bands, we would like to use all the available data to improve the group catalog which include high-quality photo-$z$ catalogs for all galaxies in the COSMOS field down to $I_{\rm AB} = 22.5$. In this section we present our method of populating the spectroscopic groups discussed in the previous chapter by photo-$z$ galaxies on a probabilistic basis.

Although there are in principal ways to detect groups in photometric galaxy samples \citep[e.g.,][]{li2008,gillis2011}, we will only use the groups detected by spectroscopic galaxies. We will not use photo-$z$ galaxies to detect new groups. Thus our resulting group sample will be missing the population of all groups in the sky that do not have more than one spectroscopic member. Inspection of Figure \ref{fig2:halo_completeness} gives information on the fraction of groups that are missed for this reason since it plots the fraction of detectable halos (i.e., those with two or more galaxies above the zCOSMOS flux limit) that actually had two or more galaxies observed spectroscopically after the incomplete spatial sampling and redshift success rates are applied.

\subsection{Assigning probabilities to photo-$z$ galaxies}\label{sec:p}

Although the photo-$z$ errors of $\sim \! 0.01(1+z)$ are impressively small by normal standards, we cannot incorporate these galaxies into the group-finding scheme directly, or even unambiguously assign them to groups in a unique and reliable way. Some group galaxies might appear at large distance from the group center in redshift space and some galaxies could be candidates for several groups. However, we can attempt to quantify the probability that galaxies are associated to a given group. This probability will depend on the distance from the group center both in the plane of the sky and in the redshift dimension. We can again use the mock catalogs to determine these probabilities, similar to their use to fine-tune the group-finding algorithm.  Additionally, the association probability may also depend on the luminosity or stellar mass of the galaxy in question. However, since this may depend on the galaxy evolution prescription in the COSMOS mock light cones and since one of our scientific goals is to use the group catalog to test such relations, we decided not use this additional information in estimating association probabilities.

Suppose we have a group at $(\alpha_{\rm gr},\delta_{\rm gr},z_{\rm gr})$ in redshift space and a nearby galaxy at $(\alpha,\delta,z)$ with a redshift error of $\delta_z$. We will parameterize the distance of the galaxy from the group by the scaled, dimensionless offsets perpendicular and parallel to the line of sight
\begin{equation}
\sigma_{r} = \frac{r}{r_{\rm gr}}\:,\qquad \sigma_{\rm z} = \frac{\left|z-z_{\rm gr}\right|}{\delta z}\:,
\end{equation}
where $r(\alpha,\delta,\alpha_{\rm gr},\delta_{\rm gr},z_{\rm gr})$ is the physical distance of the galaxy from the group center perpendicular to the line of sight and $r_{\rm gr}$ is a measure of the projected physical extension of the group. A suitable group extension parameter $r_{\rm gr}$ should ideally scale with the virial radius of the group and $(\alpha_{\rm gr},\delta_{\rm gr})$ should approach the center of the underlying DM halo. Since there are no unique estimators satisfying these requirements we will focus on different possibilities and discuss their relative strengths using the mock catalogs.

Regarding the group extension $r_{\rm gr}$, a natural estimator would be the root-mean-square (rms) extension of the spectroscopic members within the group, that is,
\begin{equation}
r_{\rm rms}(\alpha_{\rm gr},\delta_{\rm gr}) = \frac{D(z_{\rm gr})}{(1+z_{\rm gr})} \: \tilde{r}_{\rm rms}(\alpha_{\rm gr},\delta_{\rm gr}),
\end{equation}
with
\begin{equation}\label{eq:rms_tilde}
\tilde{r}_{\rm rms}(\alpha_{\rm gr},\delta_{\rm gr}) = \sqrt{\frac{1}{N} \sum_{i=1}^N \Big( \Delta\alpha_i^2 + \Delta \delta_i^2 \Big)}\:
\end{equation}
and $\Delta\alpha_i = \alpha_i-\alpha_{\rm gr}$ and $\Delta\delta_i = \delta_i-\delta_{\rm gr}$, where $(\alpha_i,\delta_i)$ is the position of the $i$th galaxy in the group and $D(z_{\rm gr})$ is the comoving distance to redshift $z_{\rm gr}$. Note that this estimator is still dependent on the choice of the group centers $(\alpha_{\rm gr},\delta_{\rm gr})$. The main drawback of this choice is its low correlation with the virial radius of the group in the mock catalogs. In fact, it proved to be very hard to estimate the virial radius from the distribution of galaxies. The second problem is based on the observation that particularly for groups with low richness $N$ the scaling $r_{\rm rms}$ can become unrealistically small because of chance orientation effects. Another approach for $r_{\rm gr}$ is the fudge radius $r_{\rm fudge}$, which has the advantage of solving both the drawbacks of $r_{\rm rms}$.

The estimators for the group centers are discussed in detail in Section \ref{sec:group_centers}. Some of the discussed estimators use also the photo-$z$ information. As a benchmark for comparison we will often use simply the average over the positions of the spectroscopic group members which will be termed ``standard centers''. On the other hand, for the final computation of association probabilities, we have used ``improved centers'' (defined in Sect.~\ref{sec:group_centers}) which are themselves based on association probabilities of photo-$z$ galaxies. So the final probabilities are obtained by an iterative procedure which, however, already converges after one iteration. 

Taking all reconstructed mock groups with 2WM to real groups, we then compute the fraction $f(\sigma_{\rm r},\sigma_{\rm z},N)$ of photo-$z$ galaxies which are members of the corresponding real group as a function of $\sigma_{\rm r}$, $\sigma_{\rm z}$, and $N$. To obtain large enough group samples for the computation of $f$, we restrict the richness dependence to just four richness classes $N = 2$, $3 \leq N \leq 4$, $5 \leq N \leq 9$, and $N \geq 10$. For each galaxy and each group, the function $f(\sigma_{\rm r},\sigma_{\rm z},N)$ is then evaluated and interpreted as the probability that this galaxy is a member of this group. Since the function $f$ was estimated using only the reconstructed groups with 2WM to real groups, it does not include the effects of deficiencies in the original detection of the spectroscopic groups (cf.~Fig.~\ref{fig2:catalog_statistics}).  In other words, $f$ is the probability that a galaxy is a member of an apparent group, defined as a certain location in ($\alpha,\delta,z$) space, to which should be multiplied the probability that the apparent group is actually real.   

\begin{figure}
	\centering
	\includegraphics[width=0.46\textwidth]{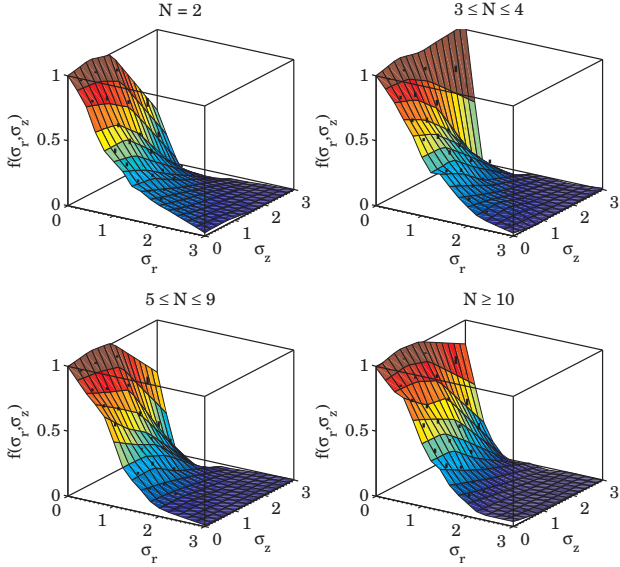}
	\caption{Fraction $f(\sigma_{\rm r},\sigma_{\rm z},N)$ of photo-$z$ galaxies to be associated to groups, where $\sigma_{\rm r}$ is based on the fudge radius and the improved centers. The surface is the mean of $f$ of the 24 mock catalogs and the error of the mean is indicated by the black bars. The function $f$ was empirically computed using the reconstructed groups exhibiting 2WM to real groups (i.e., it does not include the effect of group detection failures) and is based on $r_{\rm fudge}$ and the improved centers (see the text). For $\sigma_{\rm r} \sim 3$ or $\sigma_{\rm z} \sim 3$ the fraction $f$ is basically zero.\label{fig2:f_r_z}}
\end{figure}

The functions $f(\sigma_{\rm r},\sigma_{\rm z},N)$ are shown in Figure \ref{fig2:f_r_z} for the four richness classes.
They are all very similar and are smooth, strongly decreasing functions for increasing $\sigma_{\rm r}$ and $\sigma_{\rm z}$. Not surprisingly the probability of a galaxy being a member of the group is usually much larger than the formal integral of the redshift photo-$z$ probability distribution for that galaxy over the very small redshift interval associated with the group, which is of course the motivation for this approach.

In this scheme a galaxy can be associated to more than one group if it lies close enough to both of them, i.e., if $f(\sigma_{\rm r},\sigma_{\rm z},N)$ is non-zero in either case.  Indeed, the probabilities as computed in the previous paragraph may even sum up to more than unity. We therefore introduce a slight modification of the assigned probabilities. If a galaxy is associated to $n$ groups with probabilities $p_i$, $i = 1,\ldots,n$, we first compute the probability that it is not a member of any group
\begin{equation}
p_{\rm nongr} = \prod_{i=1}^n (1-p_i)\:.
\end{equation}
Then the probability of the galaxy to be in \emph{any} group is taken to be $1-p_{\rm nongr}$ instead of $p_{\rm tot} = \sum_{i=1}^n p_i$. Finally we just scale the probabilities by the ratio of these two, i.e.,
\begin{equation}
\tilde{p}_i = p_i \frac{1-p_{\rm nongr}}{p_{\rm tot}}\:.
\end{equation}
For the ease of notation we will just write $p_i$ instead of $\tilde{p}_i$ in the following and refer to these quantities as ``association probabilities''.

\subsection{Properties of the association probabilities}

In the following, we will study the properties of the association probabilities introduced in the previous section in terms of fidelity and completeness for different group subsamples and different choices of the group extension $r_{\rm gr}$ and group centers $(\alpha_{\rm gr},\delta_{\rm gr})$, and we will compare the distribution of probabilities in the mock catalogs to that in the actual data.

To investigate the fidelity of the association probability, we define a photo-$z$ to be ``successfully associated'' to a reconstructed mock group with a 2WM to a real group (``2WM group''), if the photo-$z$ galaxy is a member of the real group. For reconstructed mock groups with no 2WM to real groups (``non-2WM group''), i.e., the $\sim\! 30\%$ reconstructed groups which are not in the bottom layer of Figure \ref{fig2:different_associations}, the definition of a successful association is more subtle. If a non-2WM group is fragmented, a successful association is defined in the sense that the photo-$z$ is a member of the real group to which our reconstructed group is associated. In the case of an overmerged reconstructed group the photo-$z$, there is more than one real group that is associated to our reconstructed group. Here a photo-$z$ is successfully associated to the reconstructed group, if it is a member of the corresponding real group that contains the largest fraction of the members of our reconstructed group. For spurious groups, there is no corresponding real group and every photo-$z$ is regarded as failed.

\begin{figure}
	\centering
	\includegraphics[width=0.46\textwidth]{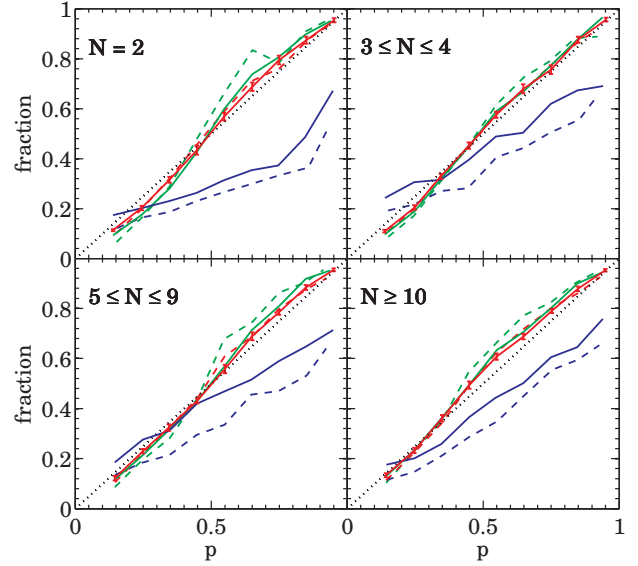}
	\caption{Fraction of correct associations as a function association probability $p$ for different richness classes. The lines show the mean of the fractions computed for each of the 24 mock catalogs and the error bars indicate the standard deviation of the mean. The solid lines correspond to estimates of $p$ based on $r_{\rm fudge}$ and the improved centers, and the dashed lines correspond to estimates based on $r_{\rm rms}$ and the standard centers. The red lines correspond to galaxies associated to reconstructed groups exhibiting a 2WM to real groups. Ideally this lines should lie on the dotted line. Statistically significant deviations are caused by galaxies which are associated to more than one group. The probabilities for such galaxies are given by the green lines. In contrast, the blue lines are for galaxies associated to group which do not exhibit a 2WM to real groups.\label{fig2:phot_p_p}}
\end{figure}

Figure \ref{fig2:phot_p_p} shows the fraction of successful associations as a function of probability $p$. The red line shows the success of associations for 2WM groups and this should be a diagonal line because the probabilities were calibrated using these groups. The green line shows the result for those galaxies in 2WM groups which have non-zero association probabilities to more than one group and also looks satisfactory. The net result for non-2WM groups is shown in blue. These curves are lower than the other curves because of the problems with group identification.  
The solid lines correspond to estimates of $p$ based on the fudge radius and the improved centers, and the dashed lines correspond to estimates based on $r_{\rm rms}$ and the standard centers.  While the choice of the group extension $r_{\rm gr}$ seems to have a negligible effect for those photo-$z$ being associated to 2WM groups (red versus green lines), the fudge radius $r_{\rm fudge}$ seems to work better for the ``failed groups''. The reason for this is that such groups have sometimes strange shapes so that $r_{\rm rms}$ is far too large which results in more (wrongly) associated galaxies than if $r_{\rm fudge}$ was used. The fudge radius $r_{\rm fudge}$ instead depends only on the richness and thus is unaffected by the shape of the group.

\begin{figure}
	\centering
	\includegraphics[width=0.46\textwidth]{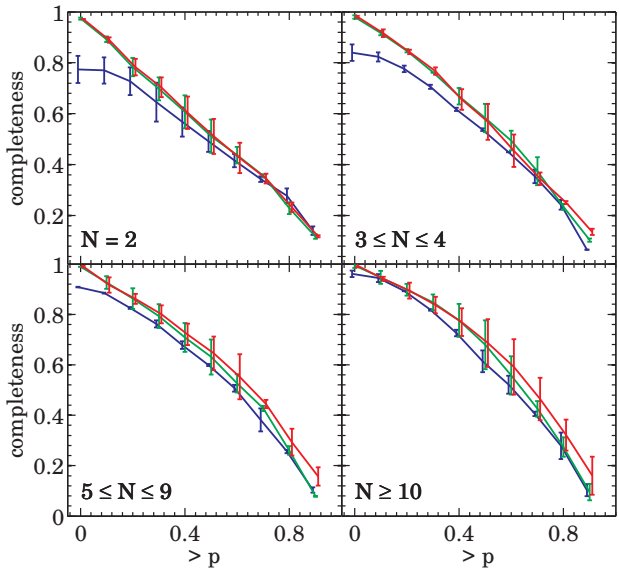}
	\caption{Completeness of the photo-$z$ group population with probabilities $>\!p$. The lines show the median of the 24 mock catalogs and the error bars exhibit the upper and lower quartiles. Only mock groups with 2WM to real groups are considered. The blue line corresponds to probabilities $p$ based on $r_{\rm rms}$, the green line to $p$ based on $f_{\rm fudge}$ and standard group centers, and the red line to $p$ based on $f_{\rm fudge}$ and on the improved centers. \label{fig2:phot_completeness}}
\end{figure}

The completeness of the group membership for all photo-$z$ galaxies above a given threshold in $p$ is shown in Figure \ref{fig2:phot_completeness}.
The blue line is for probabilities $p$ based on $r_{\rm rms}$, the green line for $p$ based on $r_{\rm fudge}$ and the standard centers, and the red line for $p$ based on $r_{\rm fudge}$ and the improved centers. The biggest difference between the blue curve (using $r_{\rm rms}$) and the other lines is at low $p$, where particularly for small groups the completeness is significantly lower than for the curves being based on $r_{\rm fudge}$. For small groups, $r_{\rm rms}$ can be an underestimate and so too few photo-$z$ galaxies are associated to such groups. This is the most significant advantage of using $r_{\rm fudge}$ instead of $r_{\rm rms}$. The difference between the choice of the group centers is most obvious at high $p$ and for large groups, where the improved centers exhibit a slight improvement. The choice of the group extension is, however, more important than the choice of the centers.

\begin{figure}
	\centering
	\includegraphics[width=0.46\textwidth]{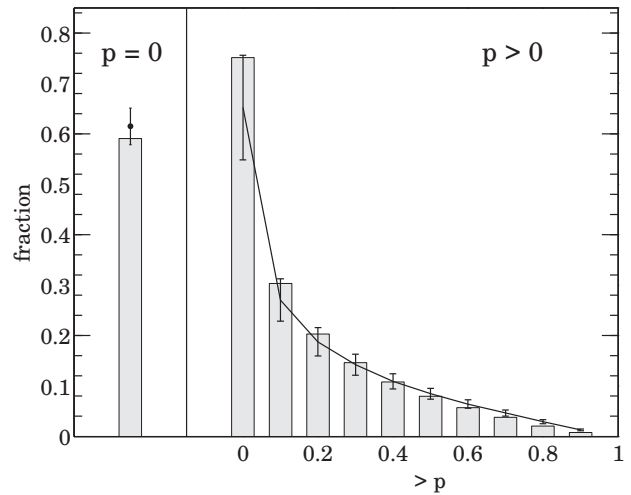}
	\caption{Fraction of photo-$z$ galaxies with an association probability $>\!p$ in the redshift range $0.1 < z < 0.8$. The histogram shows the actual data and the solid line the corresponding fraction within the mock catalogs (error bars are the standard deviation among the 24 mock catalogs). The fraction of galaxies which are not associated to any spectroscopic group (i.e., those with $p = 0$) are shown on the left. These comprise 60\% of the photo-$z$ objects.  It should be noted that the $p = 0$ and $p > 0$ fractions do not sum to unity because some galaxies have multiple $p$ values due to possible membership to different groups. The slight excess of low-probability members in the actual data is due to the larger number of small groups in the 20k sample.\label{fig2:photoz_p_histogram}}
\end{figure}

The fraction of photo-$z$ with an association probability $> \!p$ is shown in Figure \ref{fig2:photoz_p_histogram} for the actual data and for the mock catalogs. 
To allow for a meaningful comparison between galaxies with $p>0$ and galaxies with $p = 0$ we constrain the redshift range to $0.1 < z < 0.8$, where most of the groups are. About 60\% of the photo-$z$ galaxies have zero probability to be associated with any of the spectroscopic groups, while 40\% have a non-zero probability of membership of one or more groups. This fraction of possible group members drops quite fast as the $p$ threshold is increased. The slight excess of low-probability members in the actual data is due to the larger number of small groups in the 20k sample (cf.~Fig.~\ref{fig2:n_of_N}).

\begin{figure}
	\centering
	\includegraphics[width=0.5\textwidth]{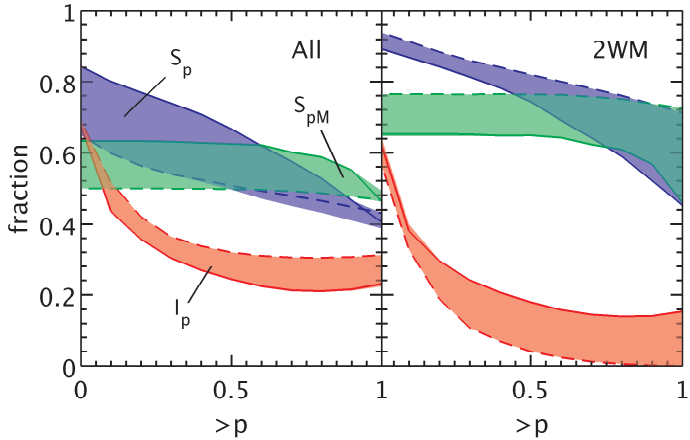}
	\caption{Average statistics for the total flux-limited galaxy sample. The left panel considers all groups in the mock catalogs and the right panel only those groups with a 2WM to real groups. The blue region shows the galaxy success rate $S_p$, the red region the interloper fraction $I_p$, and the green region the fraction of correctly assigned most massive galaxies $S_{pM}$ by picking the galaxy with the highest $p_{\rm M}$, where in each case the galaxy sample includes all galaxies with an assignment probability $>\!p$. The point $p=1$ corresponds to the spectroscopic sample. The solid lines indicates the statistics for groups with richness $N \geq 10$ and the dashed lines that for pairs. Note that a galaxy membership is here only regarded as a success if it is a member of the corresponding specific real group making these statistics rather restrictive. The poorer performance in the left hand panel is due to the issues of group detection (i.e., overmerging etc.). \label{fig2:absolute_group_membership_statistics}}
\end{figure}

The completeness and interloper fraction for the flux-limited mock group population that is obtained by including in the groups all potential members with a minimal association probability $p$ are summarized in Figure \ref{fig2:absolute_group_membership_statistics}. We show the mean completeness $S_p$ (blue region) and mean interloper fraction $I_p$ (red region) of the 24 mock catalogs, where in each mock catalog all reconstructed groups were considered (left panel). We regard only those group members as successes that are members of the corresponding real group. The point $p = 1$ corresponds to the purely spectroscopic group membership.
Note that these statistics are worse than the galaxy success rate $S_{\rm gal}$ and interloper fraction $f_{\rm I}$ shown in Figure \ref{fig2:catalog_statistics} because previously we were only concerned with whether the galaxy was a member of $any$ group.  Furthermore, here we refer to the entire flux-limited population and not only to the spectroscopic sample.

The interpretation of Figure \ref{fig2:absolute_group_membership_statistics} is as follows: looking at the claimed membership of a given reconstructed mock group, i.e., summing the spectroscopic members and all those photo-$z$ galaxies above a minimal probability threshold $p$, the new galaxy success rate $S_p$ is the number of these that are actually members of the corresponding real group divided by the total membership of this corresponding real group. This is given by the blue region of the left-hand panel which is bounded by the lines for $N = 2$ and $N \geq 10$, where $N$ refers to the observed spectroscopic richness.  The fraction of claimed galaxies that are not members of this particular group, which is the interloper fraction $I_p$ among the claimed members, is given by the corresponding red region.  As an illustration, group members of an overmerged reconstructed mock group that belong to the second real group (that is not regarded as the proper real counterpart) are regarded as failures and will increase the $I_p$ statistics (while they were not necessarily regarded as failures in the earlier $f_{\rm I}$ statistics).  If we, however, know (for reasons beyond our group catalog) that the group we are interested in is properly detected (i.e., has a 2WM to a real group), the statistics would improve to the regions in the right panel. Particularly for small groups (dashed lines), the difference will be significant owing to the uncertainties in the group detection.

\section{Applications using added photo-$z$ members}\label{sec:applications}

In this section, we perform four straightforward applications considering the potential members on the basis of their photo-$z$. We look in turn at the corrected richness, the identification of the most (stellar) massive galaxy in the group, the location of the spatial center of the group, defined as the minimum of the potential well, and finally an approach to identifying the galaxy at that center, which we define to be the central galaxy, all other group members being satellites.

Motivated by the obvious variation of the galaxy success rate and interloper fraction of the spec-$z$ group population with group-centric distance (cf.~Fig.~10 of K09), we introduced also an association probability $p$ for the spectroscopic galaxies. This will prevent spec-$z$ galaxies at the outskirts of the groups to be given a too large weight compared to their photo-$z$ group population. We assigned the probabilities in the same way as for the photo-$z$ except for the fact that we assign only probabilities to spectroscopic galaxies which were already group members and we set $\sigma_{\rm z}$ to zero, i.e., the association probability was determined only by the distance from the group center. For pairs the assigned probabilities were just set to one.

\subsection{Corrected richness}\label{sec:corrected_richness}

A straightforward application of the association probability $p$ is to estimate the corrected richness $N_{\rm corr}$ of the groups above the flux limit, i.e., the total richness the groups would have if we knew all their real members down to the flux limit of the survey, by summing up all probabilities of the group members (spec-$z$ and photo-$z$). Not surprisingly, the estimated corrected richness is on average unbiased with respect to the real corrected richness for all observed spectroscopic richness classes $N$, because this was used in establishing the probabilities. It exhibits a scatter of about 30\%, weakly depending on $N$.

The corrected richness could also be estimated by considering the SSR and RSR (see Sect.~\ref{sec:data}) at the positions of the spec-$z$ group members. However, the resulting corrected richnesses are biased for groups with $N \lesssim 4$ by being about 40\% too high and also the scatter is larger being about 50\%. The reason for this bias for small groups is a selection effect. Since the observed richness $N$ is the result of a Poisson sampling process when assigning the slits to the targets, it has an intrinsic scatter for a given SSR and RSR. If, however, $N$ drops below 2, the group cannot be observed and is lost, while for the scatter toward high $N$ there is no such limit.

We conclude that the photo-$z$ are useful for obtaining unbiased estimates of the corrected richness for all groups. We can, of course, also estimate the corrected richness $N_{\rm corr}(M_{\rm B,lim})$ with respect to a given absolute magnitude limit $M_{\rm B,lim}$ (cf.~Sect.~4.2 of K09).

\subsection{Identifying the most massive galaxy of the group}\label{sec:p_M}

We introduce the probability $p_{\rm M}$ of a galaxy to be the most massive (in terms of stellar mass) of a given group. This is done by sorting all the members---spectroscopic as well as photometric---in descending order of mass such that $M_{i-1}\geq M_i$ for $i \in 2,\ldots,N_{\rm tot}$, where $N_{\rm tot}$ is the number of spectroscopic and photometric members. The probability $[p_{\rm M}]_i$ of a given galaxy is the probability that it is the most massive galaxy in the group, which will depend on both its own probability of membership, $p_i$, and the probabilities of non-membership of higher ranked galaxies, i.e., for the first-ranked galaxy, $[p_{\rm M}]_1 = p_1$, and for the remainder,
\begin{equation}
[p_{\rm M}]_i = p_i \prod_{j=1}^{i-1} \left(1-p_j\right)\:,\quad i  \in 2,\ldots,N_{\rm tot}\:.
\end{equation}

\begin{figure}
	\centering
	\includegraphics[width=0.46\textwidth]{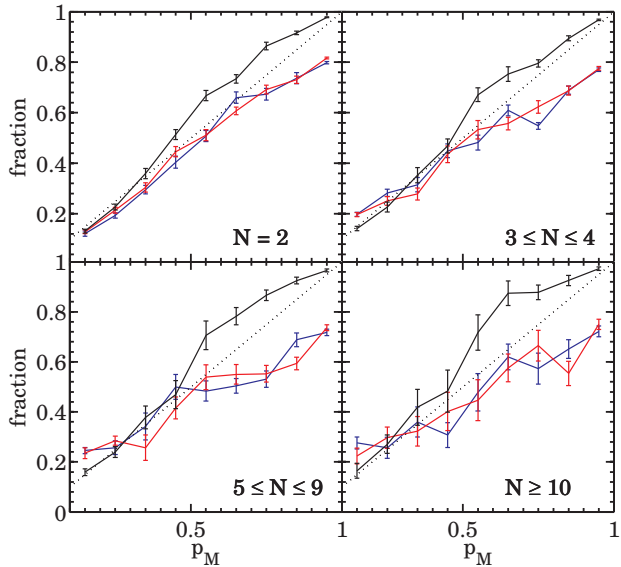}
	\caption{Fraction of correct most massive galaxies within reconstructed 2WM mock groups as a function of $p_{\rm M}$. Each panel is for a different richness class as indicated. The lines show the mean of the fractions computed for each of the 24 mock catalogs and the error bars indicate the error of the mean. The red curve corresponds to $p_{\rm M}$ based on $r_{\rm fudge}$ and the sophisticated group centers, and the blue curve corresponds to $p_{\rm M}$ based on $r_{\rm rms}$ and the standard centers. The black line corresponds to the former case, but does not include observational errors in stellar mass of 0.2 dex.\label{fig2:p_M_association}}
\end{figure}

Figure \ref{fig2:p_M_association} compares the $p_{\rm M}$ to the empirical fraction of correctly identified most massive galaxies within the mock catalogs.
Ideally, this would be the dotted diagonal line, in that galaxies with some value of $p_{\rm M}$ should be the real most massive galaxies in a fraction $p_{\rm M}$ of cases. The red curve uses association probabilities $p$ based on $r_{\rm fudge}$ and the blue curve those based on $r_{\rm rms}$. The black curve is based on $r_{\rm fudge}$, but does not include observational errors in stellar mass determination, which are included at the level of 0.2 dex in the red and blue curves. The conclusion is that the basic scheme works (as would be expected) but that mass estimation uncertainties will be significant. While it makes no substantial difference whether the association probabilities $p$ for the computation of $p_{\rm M}$ are based on $r_{\rm rms}$ or $r_{\rm fudge}$, the uncertainty in the stellar mass of 0.2 dex causes the $p_{\rm M}$ to be underestimated for large $p_{\rm M}$. Nevertheless there is a strong correlation between $p_{\rm M}$ and the fraction of cases in which the galaxy under consideration is the real most massive galaxy of the group. For a cut, for instance, of $p_{\rm M} > 0.7$, the true probability is still higher than $50\%$, i.e., such a galaxy has a bigger chance of being the most massive galaxy than all the other candidates in its group put together. For a proper interpretation of $p_{\rm M}$ it is, however, important to keep this effect of the mass uncertainty in mind.

\begin{figure}
	\centering
	\includegraphics[width=0.46\textwidth]{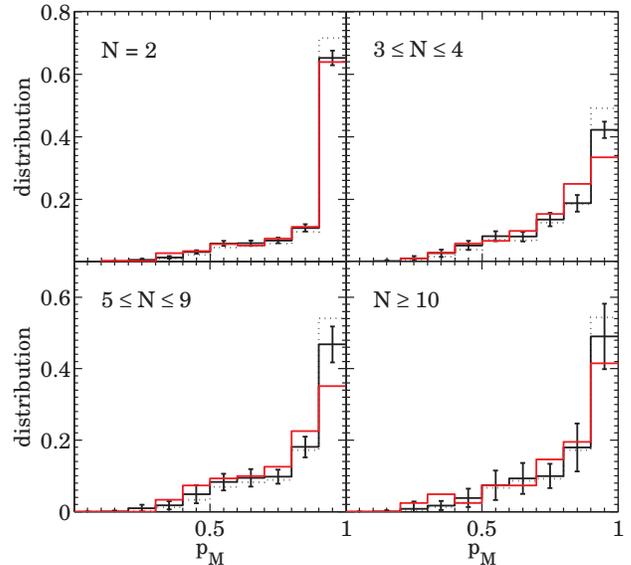}
	\caption{Distribution of the probabilities $p_{\rm M}$ of those galaxies with the highest $p_{\rm M}$ in their groups. These galaxies can be either spectroscopic or photometric. Each panel corresponds to a different richness class. The black histogram corresponds to the mean distribution of the 24 mock catalogs and the error bars indicate the standard deviation. The dotted histogram considers only the mock groups having a 2WM to real groups. The red histogram shows the histogram for the actual zCOSMOS 20k groups. It is obvious that most groups have a clearly identified candidate for being the most massive galaxy. \label{fig2:p_M_histogram}}
\end{figure}

In assessing the usefulness of this scheme, this figure should be combined with Figure \ref{fig2:p_M_histogram} which shows the distribution of $p_{\rm M}$ as a function of richness class, for both the mock catalogs and for the actual 20k data.
Note that for each richness class, the distribution of $p_{\rm M}$ of those galaxies with the highest $p_{\rm M}$ in their groups is a steep function of  $p_{\rm M}$. This tells us that most groups have a clear candidate for being the most massive galaxy. The actual 20k sample (red histogram) follows fairly well the histogram for the mock catalogs (black solid) except maybe for the largest $p_{\rm M}$. Despite the uncertainty in stellar mass, $p_{\rm M}$ is a very useful concept and works reasonably well for the actual data.

It should be noted that in Figure \ref{fig2:p_M_association} the $p_{\rm M}$ for the unperturbed stellar masses slightly underestimates the true probability for $p_{\rm M} \gtrsim 0.5$ as measured by the fraction of such galaxies that really are the most massive members of their groups, i.e., the black line lies slightly above the dotted diagonal line. This is due to the fact that the association probabilities $p$ were derived irrespective of the mass or luminosity of the galaxies. If massive galaxies are more likely to be in groups, then this process will have underestimated the $p_i$ and thus $p_{\rm M}$ for these more massive galaxies, producing the small offset observed in Figure \ref{fig2:p_M_association}. 

The average success rate $S_{pM}$ of detecting the most massive galaxy within the reconstructed mock groups as a function of a probability $p$ threshold is shown in Figure \ref{fig2:absolute_group_membership_statistics}. For high richness groups, this success rate increases by about 10\%-20\% when we include photo-$z$ galaxies, while it is relatively constant for spec-$z$ pairs. So the inclusion of the photo-$z$ galaxies has a rather small effect on $S_{pM}$. In fact, for 87\% of all groups the galaxy with the largest $p_{\rm M}$ has a spectroscopic redshift (for the mock catalogs this number is $84\% \pm 2$\%).  It might be thought that this ratio should be equal to the average spatial completeness of the survey, since this determined the chance that a given galaxy is observed spectroscopically.  This will be the case for very rich groups, which would be recognized regardless of the statistical fluctuations of spatial sampling. For poorer groups, there is however a selection effect in that those with higher spectroscopic sampling will be more likely to be recognized as a group.  Indeed, for ``real'' pairs, both members must have been observed spectroscopically for the group to be recognized, and the most massive galaxy will therefore always be a spectroscopic galaxy. 

Does this relatively modest gain mean that it is not worth bothering with the photo-$z$? The answer is no. First, we show in the next section that there are significant gains in finding the spatial center of the group.  Second, the inclusion of photo-$z$ objects dramatically reduces the number of galaxies that are incorrectly identified as the most massive in the richer groups. These may be among the most interesting from a galaxy evolution point of view. It should also be noted that for these statistics the identification of the most massive galaxy is only regarded as a success if it is the most massive galaxy of the specific group we think it is a member of. Selecting a galaxy that is the real most massive galaxy of another group (even one that has been detected) is considered here as a failure. This is a rather restrictive perspective and depending on the application it may be sufficient to just know whether a certain galaxy is the most massive of \emph{any} group (cf.~Sect.~\ref{sec:central_galaxy}).

\subsection{Locating the spatial group center}\label{sec:group_centers}

Another immediate application of the association probabilities $p$ is to estimate the centers of the groups. By group center we mean the center of the corresponding DM halo which is defined by the position of the deepest point in the gravitational potential well. In the Millennium simulations this position is given by the most bound particle within a halo and is also, by construction, the position of the ``central galaxy'' in the halo.

With the aid of the mock catalogs we can test several estimators $E$ for the group center and compare their relative accuracy. Some of these estimators are based on the areas of the Voronoi cells of the projected group galaxy positions \citep[see also][]{presotto2012}. To compute these areas we project a group to the plane perpendicular to the line of sight and perform a two-dimensional Voronoi tessellation considering only the group galaxies (either only spec-$z$ or both spec-$z$ and photo-$z$) and the spectroscopic field galaxies surrounding the group to prevent the areas of the Voronoi cells at the outskirts of the group to become infinite. We expect the size of the Voronoi areas to be smaller on average toward the center of the groups.

In the following, using the mock catalogs we will test 10 different estimators, $E_1$ to $E_{10}$, to identify the group centers. The estimators $E_1$ to $E_4$ are depending on the spectroscopic information only:
\begin{itemize}
	\item[$E_1$:] mean of the positions of the spectroscopic members;
	
	\item[$E_2$:] stellar mass weighted mean of the positions of the spectroscopic members;
	
	\item[$E_3$:] inverse Voronoi area weighted mean of the positions of the spectroscopic members;
	
	\item[$E_4$:] stellar mass and inverse projected Voronoi area weighted mean of the positions of the spectroscopic members.
\end{itemize}
The estimators $E_5$ to $E_8$ include also the information from the photometric galaxies. They are basically identical to the former estimators, but that each galaxy---spec-$z$ as well as photo-$z$---is additionally weighted by their association probability $p$:
\begin{itemize}
	\item[$E_5$:] probability weighted mean of the positions of all group members;
	
	\item[$E_6$:] probability and stellar mass weighted mean of the positions of all group members;
	
	\item[$E_7$:] probability and inverse Voronoi area weighted mean of the positions of all group members;
	
	\item[$E_8$:] probability, stellar mass, and inverse Voronoi area weighted mean of the positions of all group members.
\end{itemize}
The estimators $E_9$ and $E_{10}$ are not any more based on the (weighted) mean of the positions of the group members, but attempt to find directly the central galaxies of the groups. They are defined by selecting for each group the galaxy with the largest ratio $R$, as follows:
\begin{itemize}
	\item[$E_{9\ }$:] location of the galaxy with the largest $R = p /A$;
	
	\item[$E_{10}$:] location of the galaxy with the largest $R = p M_{\ast}/A$.
\end{itemize}
Here, $M_{\ast}$ is the stellar mass and $A$ the Voronoi area of the galaxies. Note that all estimators which are based on the Voronoi area $A$ are not necessarily defined for all groups since $A$ might happen to be infinite for the members of small isolated groups near to the border of the survey.

\begin{figure*}
	\centering
	\includegraphics[width=0.85\textwidth]{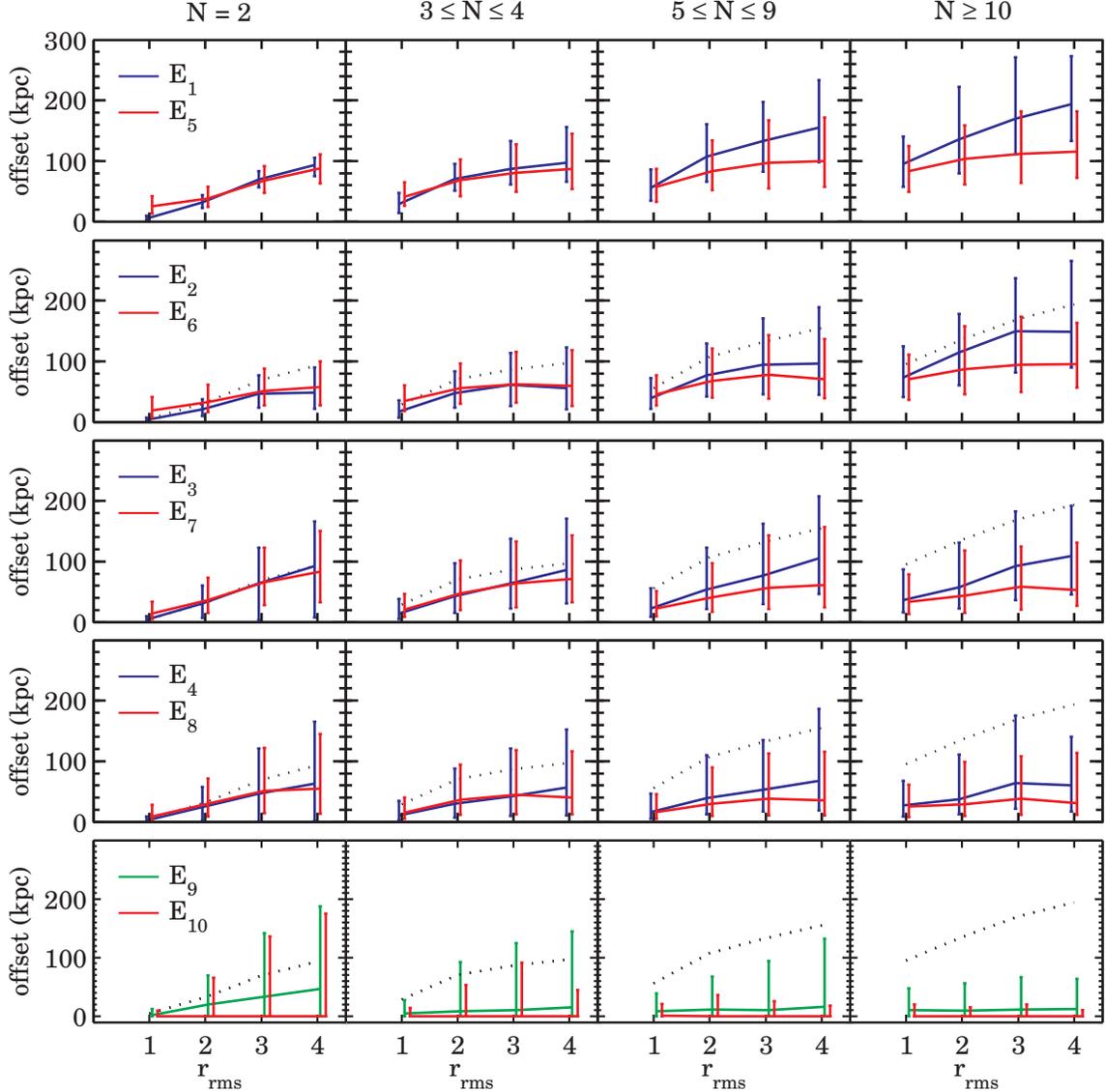}
	\caption{Projected physical offset between the estimated group centers and the true group centers in the mock catalogs. The lines show the median offsets of all reconstructed 2WM groups within the 24 mock catalogs, and the error bars indicate the upper and lower quartiles. The $x$-axis plots the four quartiles for the apparent group extension $\tilde r_{\rm rms}$ (see Tab.~\ref{tab2:r_rms_quartiles}). The estimators are indicated for each row in the left panel and the richness class increases toward right. Blue lines contain only spec-$z$ information, and red and green lines contain spec-$z$ and photo-$z$ information. For comparison, $E_1$ is shown in all panels as dotted line.\label{fig2:centers_quartile}}
\end{figure*}

The average physical offsets between the estimated mock group centers and the real group centers are shown in Figure \ref{fig2:centers_quartile} for different richness classes and for different apparent group extensions $\tilde r_{\rm rms}$ (see Eq.~(\ref{eq:rms_tilde})).
The dependence on $\tilde r_{\rm rms}$ is considered by dividing each group population within a richness class into the four quartiles of its distribution in $\tilde r_{\rm rms}$. The values of the boundaries dividing the four quartiles are given in Table \ref{tab2:r_rms_quartiles}. The goodness of the different estimators $E_1$ to $E_{10}$ strongly depends on both the extension and the richness of the group. There can be several statements made.
\begin{deluxetable}{cccc}
\tablewidth{0pt}
\tablecaption{Borders dividing the four quartiles of the distribution of $\tilde r_{\rm rms}$ for group populations of different richness classes}
\tablehead{
	\colhead{} &
  \colhead{1/2-quartile} &
  \colhead{2/3-quartile} &
  \colhead{3/4-quartile}
  }
\startdata

$N = 2$           & 0.0011 & $0.0034$ & $0.0055$  \\
$3 \leq N \leq 4$ & 0.0042 & $0.0064$ & $0.0089$  \\
$5 \leq N \leq 9$ & 0.0079 & $0.0109$ & $0.0150$  \\
$N \geq 10$       & 0.0140 & $0.0193$ & $0.0273$    
\enddata

\tablecomments{The values are given in degrees.}

\label{tab2:r_rms_quartiles}
\end{deluxetable}

\begin{itemize}
	\item The smaller the group extension, that is, the smaller the group appears projected on the sky, the smaller the offset from the true group center.
	
	\item The larger the observed richness of the group the more effective is weighting the galaxies by stellar mass or the inverse of the Voronoi area.
	
	\item For group with $N \lesssim 5$ weighting by stellar mass is more effective than weighting by the inverse of the Voronoi area and reversed for $N \gtrsim 5$.
	
	\item Weighting by both stellar mass and the inverse of the Voronoi area is superior to weighting by either of these alone for all richness classes except for pairs.
	
	\item The larger $N$ and the more extended the group, the more effective is the consideration of the photo-$z$ galaxies. For $N \lesssim 5$ there is hardly any gain from using the photo-$z$ information, whereas for $N \gtrsim 5$ there is a clear gain for all estimators particularly for extended groups.
	
	\item The size of the error bars suggests that the most robust estimator for pairs is the simple geometrical mean between the two galaxies.
	
	\item For all groups except of pairs, by far the best estimator is $E_{10}$. For groups with $N \geq 5$ for at least half of the groups of any extension the central galaxy (not necessarily the most massive, see below) is correctly identified (i.e., offset equals zero) and for three quarter of the groups the offset from the true group center is less than about 20 kpc. Compared to the typical extension of a group of the order of half an Mpc, this is an extremely good result.
			
\end{itemize}

However, regarding $E_{10}$ we have to be careful since the mock catalogs by definition have a massive galaxy at the centers of their groups. This ``central galaxy paradigm'' is still under investigation (see e.g., \citealt{skibba2011}). Also note that in the mock catalogs the central galaxy of a group is not always the most massive, but in about 20\%-25\% of all real mock groups---depending weakly on $N$---there is a more massive galaxy within the magnitude-limited group population.

Also note that 50\%-60\% of the galaxies selected by $E_{10}$ are also the galaxies with the highest probability $p_{\rm M}$ within the group. Although the two concepts are similar, they are not equal. In particular, $p_{\rm M}$ can be assigned to each galaxy in a group and also yields a quantitative measure of fidelity rather then just selecting a particular galaxy without giving any information ``how good'' this selection is. Moreover $p_{\rm M}$ is totally independent on the assumption where massive galaxies preferentially reside within a group.

Given the results in Figure \ref{fig2:centers_quartile} we have chosen the ``improved group centers'', in contrast to the ``standard group centers'' being $E_1$, as follows. For groups with $N = 2$ we kept the centers to be $E_1$, for $3 \leq N \leq 4$ we took $E_3$ if available (93\% of cases) and otherwise $E_1$, and for $N \geq 5$ we took $E_7$ (always available). So only for $N \geq 5$ we have used information from photo-$z$ and in neither case we have used information from stellar mass. Inspection of Figure \ref{fig2:centers_quartile} shows that the improved group centers should exhibit offsets $\lesssim \! 100$ kpc for basically all richness classes and group extensions.

The effect on the group center estimates $E_1$-$E_{10}$, if we use association probabilities $p$ based on the improved centers instead of the standard centers, is strongest for $E_5$, especially for rich and extended groups. The differences in the offsets are, however, never larger than $\sim\! 30 \%$ and for $E_7$ they are basically negligible. This shows that the iterative process for deriving the assocation probabilities indeed converges after one iteration.

\subsection{Separating central and satellite galaxies}\label{sec:central_galaxy}

As mentioned in the previous section, the central galaxies, which we defined to be those galaxies located at the minimum of the gravitational potential, are not necessarily the most massive galaxies within the halos.  However, in terms of evolutionary processes, it is likely that it is the location in the potential well that is most relevant, and so in the following we will discuss how well we can differentiate between the central galaxies and the remaining, so-called satellite galaxies. For both centrals and satellites we will differentiate between simply knowing whether a galaxy is a central or a satellite, and the more stringent case of additionally knowing which group or halo it is the central or satellite of.   

\begin{figure*}
	\centering
	\includegraphics[width=0.76\textwidth]{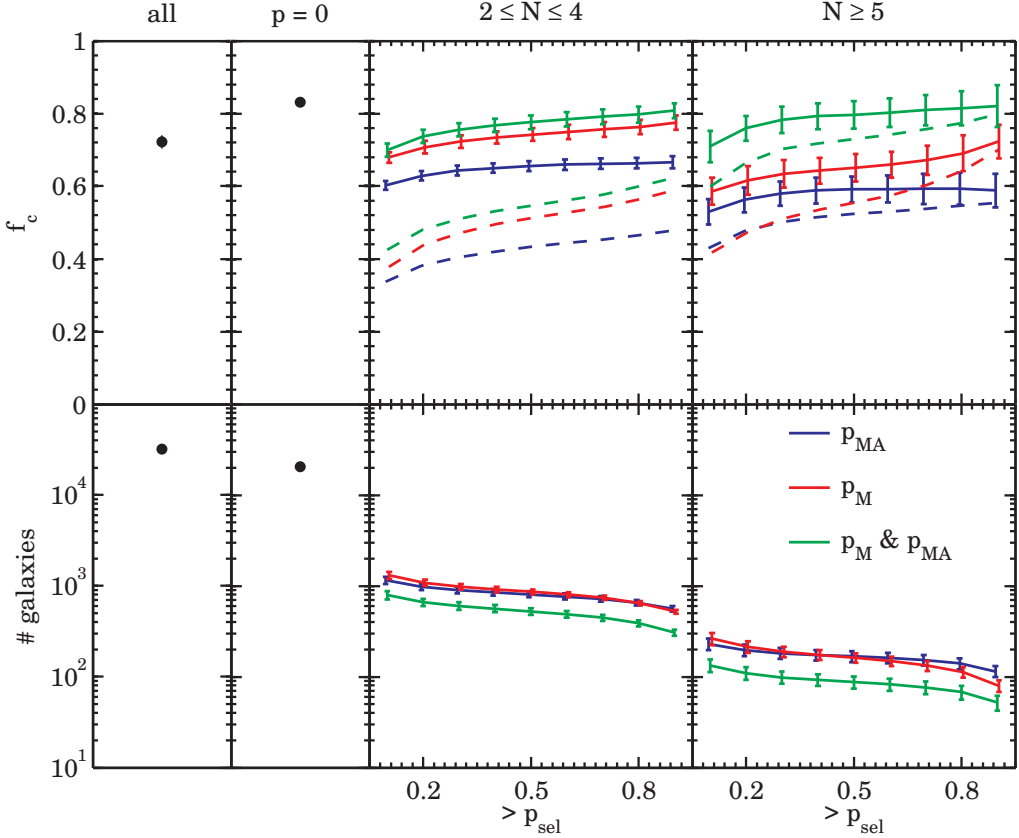}
	\caption{Fraction of centrals $f_{\rm c}$ for different mock galaxy samples in the redshift range $0.1 < z < 0.8$. \emph{Upper panels:} the left panel shows the fraction of centrals for the total flux-limited sample, and the second from the left the fraction for those galaxies (spec-$z$ and photo-$z$) which are not associated to any group (i.e., $p = 0$). The third and the forth panel from the left show the fraction of central galaxies within our mock group population for two different richness classes, where the selection probability $p_{\rm sel}$ is $p_{\rm MA}$ for the blue lines, $p_{\rm M}$ for the red lines, and the intersection of the two for the green lines. The solid lines correspond to the fraction of galaxies being the centrals of \emph{any} group, while the dashed lines correspond to the fraction of correctly identified centrals of the corresponding specific real group. \emph{Lower panels:} the number of mock galaxies in the selected samples. In all panels, the error bars indicate the standard deviation among the 24 mock catalogs. For the points in the left two panels, the error bars are smaller than the size of the points. Note that only galaxies with $p_{\rm sel}>0.1$ are shown. The fractions and numbers for galaxies with $0 < p_{\rm sel}<0.1$ deviate much from the relatively constant curves shown here (cf.~Fig.~\ref{fig2:satellite_fraction}). Note that the difference between the solid and the dashed lines for groups with $2 \leq N \leq 5$ comes mainly from the uncertainty in the detection of pairs, and note that the numbers of centrals in the actual data are similar to those in the mock catalogs (cf.~Tab.~\ref{tab2:central_satellite_fraction}). \label{fig2:central_fraction}}
\end{figure*}

\begin{figure}
	\centering
	\includegraphics[width=0.46\textwidth]{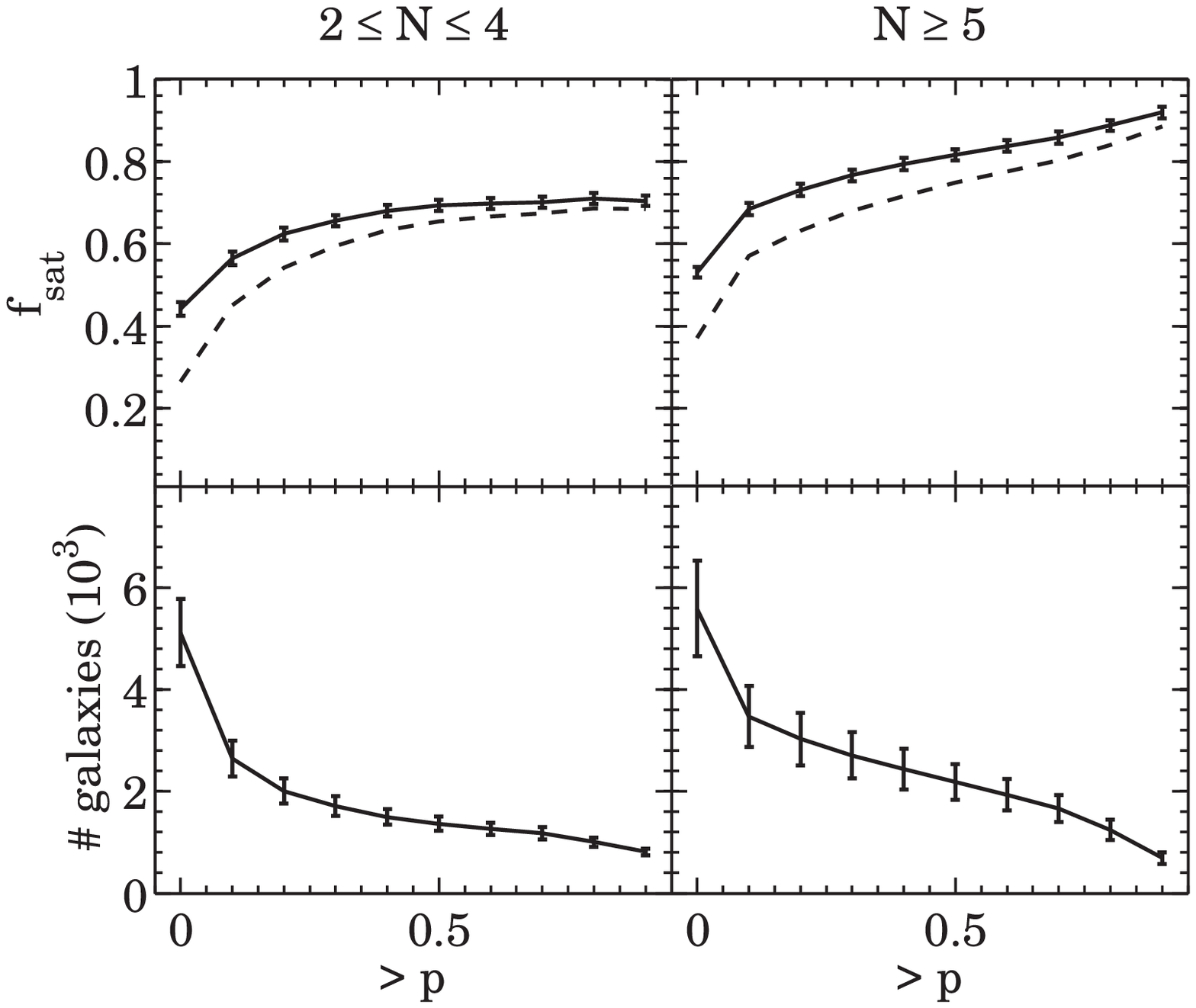}
	\caption{Fraction of satellites $f_{\rm sat}$ in the mock group population selected by $p_{\rm M} < 0.1$ and $p_{\rm MA} < 0.1$ as a function of association probability $>\! p$. The two panels correspond to the richness classes as indicated and the galaxy samples are restricted to the redshift range $0.1 < z < 0.8$. \emph{Upper panels:} the solid lines refer to the fractions of selected galaxies to be satellites of \emph{any} group, while the dashed lines correspond to the fractions of selected galaxies to be satellites of the corresponding specific real group. \emph{Lower panels:} the number of mock galaxies in the selected samples. In any panel, the error bars show the standard deviation among the 24 mock catalogs. Note that the numbers of satellites in the actual data are similar to those in the mock catalogs (cf.~Tab.~\ref{tab2:central_satellite_fraction}).\label{fig2:satellite_fraction}}
\end{figure}

Can we add spatial information to $p_{\rm M}$? Motivated by the performance of the group center estimator $E_{10}$ (see Fig.~\ref{fig2:centers_quartile}), we introduce another probability $p_{\rm MA}$ which is computed similarly to $p_{\rm M}$, but instead of ranking the group galaxies by their stellar mass $M_\ast$ we rank them by $M_\ast/A$ with $A$ being the area of the projected Voronoi cell of the galaxy and thus includes directly information on the local galaxy density.  It should however be noted that $p_{\rm M}$ already includes some positional information because of the radial dependence of the group membership probability $p$.  In the following we will discuss how the fraction of centrals $f_{\rm c}$ and satellites $f_{\rm sat}$ varies across different galaxy samples that are selected in terms of $p$, $p_{\rm M}$, $p_{\rm MA}$, and $N$.\footnote{Owing to the multiple group associations of some photo-$z$ galaxies, a selection by $p$, $p_{\rm M}$ and $p_{\rm MA}$ does, in general, not lead to a sample of galaxies with unique group membership. If needed, we resolve this degeneracy by taking for each galaxy that has multiple associations to groups that association with the highest $p$.} The results are summarized in Figures \ref{fig2:central_fraction} and \ref{fig2:satellite_fraction} and Table \ref{tab2:central_satellite_fraction}.  To allow for a sensible comparison between the number of central and satellites in the group and non-group galaxy population, we restrict the redshift range to $0.1 < z < 0.8$.

The fraction of galaxies $f_{\rm c}$ which are the central galaxies of their DM halo is shown in Figure \ref{fig2:central_fraction} for different samples of galaxies from the mock catalogs.
It should be noted that 72\% of galaxies in the overall flux-limited galaxy sample, selected irrespective of any group membership, are central galaxies (left panel).  If those galaxies (with either spec-$z$ or photo-$z$) which are associated to groups (i.e., which have $p>0$) are excluded, then this fraction rises to 83\% (second panel from the left).   So, if a large sample of central galaxies is needed, irrespective of the halos in which they reside, then simply selecting the non-group galaxies will already produce a rather pure sample, albeit one biased to lower mass halos.

However, if we want a sample of centrals extending up into the range of halo masses of our groups, we can still produce fairly pure samples by making a cut in either $p_{\rm M}$ or $p_{\rm MA}$, or both, as shown in the third and fourth panels from the left. Interestingly $p_{\rm MA}$ actually does worse than $p_{\rm M}$, but making a cut in $p_{\rm M}$ and $p_{\rm MA}$ simultaneously produces a very pure sample of centrals at the cost of numbers (green curves). For instance, by making the simultaneous cut $p_{\rm M} > 0.5$ and $p_{\rm MA} >0.5$ in groups with $N \geq 5$, we obtain a sample of about 100 centrals that are pure at the level of 80\%, in the sense that 80\% of the galaxies are indeed centrals.  However, 10\% of these are actually centrals of a different halo than that identified in the reconstructed group catalog, and so the purity defined in terms of being the central of a correctly identified group (i.e., with a 2WM) reduces to about 70\% (dashed lines, see Tab.~\ref{tab2:central_satellite_fraction}). 

In Figure \ref{fig2:central_fraction} we show the central fraction $f_{\rm c}$ only for those galaxies with a selection probability $p_{\rm sel} > 0.1$, where $p_{\rm sel}$ was either $p_{\rm M}$ or $p_{\rm MA}$ or the intersection of the two. For the remaining galaxies, the fraction $f_{\rm c}$ is much lower than for the sample with $p_{\rm sel} > 0.1$, so this sample naturally consists mostly of satellites. The fraction of satellites $f_{\rm sat}$ in this sample is shown in Figure \ref{fig2:satellite_fraction} as a function of additional selection by the association probability $p$.
While the curves in Figure \ref{fig2:central_fraction} do basically not depend on an additional selection in $p$, the fractions of satellites in Figure \ref{fig2:satellite_fraction} are sensitive to a lower limit in $p$. On the other hand, the choice of $p_{\rm sel}$ is negligible for the fraction of satellites. 

The interpretation of Figure \ref{fig2:satellite_fraction} is relatively straightforward: if a galaxy has a very low $p_{\rm sel}$ but simultaneously a high association probability $p$, it should be a satellite. For a probability selection of $p > 0.5$ for groups with $N \geq 5$, we obtain a sample of $\sim\! 2000$ galaxies of which about 80\% are indeed satellites of a group (solid lines), and 75\% are satellites in the specific groups that we think they are in (dashed lines, see Tab.~\ref{tab2:central_satellite_fraction}).
\begin{deluxetable}{crcccccc}
\tablewidth{0pt}
\tablecaption{Fractions of centrals and satellites in different galaxy samples in the redshift range $0.1 < z < 0.8$}
\tablehead{
	\colhead{Sample} &
  \colhead{Number\tablenotemark{a}} &
  \colhead{} &
  \multicolumn{2}{c}{$f_{\rm c}$} &
  \colhead{} &
  \multicolumn{2}{c}{$f_{\rm sat}$}\\
  \cline{4-5} \cline{7-8} 
  \colhead{} &
  \colhead{} &
   \colhead{} &
  \colhead{Any\tablenotemark{b}} &
  \colhead{Spec.\tablenotemark{c}} &
  \colhead{} &
  \colhead{Any\tablenotemark{b}} &
  \colhead{Spec.\tablenotemark{c}} 
  }
\startdata
All          & 30231 & & 0.72 & -- & & 0.28 & -- \\
$p = 0$ & 19161 & & 0.83& -- & &0.17& -- \\

\hline
\sidehead{Group galaxies selected by $p_{\rm M} > 0.5$ and $p_{\rm MA} > 0.5$}

$N = 2$                & 453 & & 0.77 & 0.50 & &--& --\\
$3 \leq N \leq 4$  & 183 & & 0.80 & 0.65 & &--&--\\
$5 \leq N \leq 9$  & 72   & & 0.78 & 0.70 & &--&--\\
$N \geq 10$         & 24   & &  0.79 & 0.75 & &--&--\\
  
\hline
\sidehead{Group galaxies selected by $p > 0.5$, $p_{\rm M} < 0.1$, and $p_{\rm MA} < 0.1$}
$N = 2$                & 759  & &--&--& & 0.65 & 0.61\\
$3 \leq N \leq 4$  & 794  & &--&--& & 0.74 & 0.70\\
$5 \leq N \leq 9$  & 976  & &--&--& & 0.79 & 0.73\\
$N \geq 10$         & 956 & &--&--& & 0.84 & 0.76
\enddata

\tablenotetext{a}{Number of galaxies in the corresponding actual data samples. For the mock catalogs see Figs.~\ref{fig2:central_fraction} and \ref{fig2:satellite_fraction}.}
\tablenotetext{b}{Statistics for centrals and satellites irrespective of their specific group memberships.}
\tablenotetext{c}{Statistics for centrals and satellites for residing in the specific groups we think they are members of.}

\label{tab2:central_satellite_fraction}
\end{deluxetable}

We may want to simply try to classify all galaxies as either central or satellites. That is, one does not just produce subsamples of centrals or satellites with high purity, but the samples of centrals and satellites are complementary and add up to the flux-limited sample. For any such division we can compute the completeness as well as the purity for either centrals and satellites, where we are not interested in their specific group membership. Note that the purity for centrals and satellites are just given by $f_{\rm c}$ and $f_{\rm sat}$, respectively, for the corresponding sample. For both samples, the purity and completeness are anti-correlated, such that a high purity implies a low completeness and vice versa. Additionally, the purity of satellites will be anti-correlated with the completeness of centrals and vice versa. This is similar to optimizing the group-finding parameters to obtain an optimal group catalog (cf.~Fig.~4 of K09). One has to tune the parameters $p$, $p_{\rm M}$, etc., to find the best compromise between the completeness and purity of either sample. A sensible compromise for producing the satellite sample is selecting galaxies by $p > 0.1$, $p_{\rm M} < 0.5$, and $p_{\rm MA}<0.5$, while all non-satellite galaxies constitute centrals (see Tab.~\ref{tab2:central_satellite_purity_completeness}). This yields a completeness and purity of centrals of 89\% and 81\%, respectively, and a completeness and purity of satellites of 45\% and 62\%, respectively.
\begin{deluxetable}{lccccc}
\tablewidth{0pt}
\tablecaption{Completeness and purity for complementary samples of centrals and satellites in the redshift range $0.1 < z < 0.8$}
\tablehead{
	\colhead{Sample\tablenotemark{a}} &
	\multicolumn{2}{c}{Centrals\tablenotemark{b}} &
	 \colhead{} &
	\multicolumn{2}{c}{Satellites\tablenotemark{c}} 
\\
  \cline{2-3} \cline{5-6} 
  \colhead{} &
  \colhead{Compl.} &
   \colhead{Purity} &
    \colhead{} &
  \colhead{Compl.} &
   \colhead{Purity} 
  }
\startdata

Spec-$z$ \& Photo-$z$		& 0.89 & 0.81 & & 0.45 & 0.62\\
Spec-$z$ only					& 0.93 & 0.84 & & 0.54 & 0.74
\enddata

\tablenotetext{a}{All galaxies are subjected to a binary central-satellite classification.}
\tablenotetext{b}{Centrals are given by all non-satellite galaxies.}
\tablenotetext{c}{Satellites are selected by  $p>0.1$, $p_{\rm M} < 0.5$, and $p_{\rm MA} < 0.5$.}

\label{tab2:central_satellite_purity_completeness}
\end{deluxetable}

Since the spectroscopic 20k sample constitutes basically an unbiased subsample of the total flux-limited sample, we can restrict our study of centrals and satellites to the spectroscopic galaxies (once we have used photo-$z$ objects to help classify them). In this case, the completeness and purity for centrals are 93\% and 84\%, respectively, and the completeness and purity for satellites 54\% and 74\%, respectively. Note that especially the statistics of the satellites have improved because the group membership is much better constrained for the spec-$z$ sample.

The conclusion of this section is that by applying a selection of galaxies in $p$, $p_{\rm M}$, $p_{\rm MA}$, and $N$ we can produce samples of centrals and satellites of varying purity and size. As expected, the size of the sample decreases with increasing demands on purity. Different levels of purity can also be obtained at the cost of biases in halo mass. For instance, a very large set of highly pure centrals (83\% pure) is obtained by excluding all galaxies that can possibly be associated with any detected group, but this obviously then excludes all centrals in the more massive halos we have detected. Dividing all objects in the flux-limited sample into centrals and satellites yields a set of centrals that is 81\% pure and a set of satellites that is 62\% pure. As with most other aspects of identifying groups at high redshift, the actual construction of samples must be carefully considered in the light of the scientific requirements.

\section{Discussion}\label{sec:discussion}

In this section we summarize the main properties of our 20k group catalog and comment on the general difficulties of producing high-quality group catalogs.

The catalog that we have presented in this paper contains almost 1500 groups of which  $\sim\! 570$ host three or more spectroscopic galaxies. Based on detailed analyses using realistic simulated mock group catalogs, about 75\% of the groups with three or more members should be real in the sense that they exhibit a one-to-one correspondence (i.e., 2WM) to real groups. The remainder are either fragmented, overmerged, or entirely spurious. The overall purities and completenesses for these groups (even relative to only ``detectable groups'' and if we do not care about the nature of the group, i.e., 1WM) are about 83\%. For groups that host only two spectroscopic galaxies, the statistics are even slightly worse. Fortunately, for groups with more than two spec-$z$ members these statistics are basically independent of the observed spectroscopic richness $N$ over a broad range of redshift and the number of groups as a function of $N$ should be an unbiased tracer for the number of real groups.

Given the work involved, this overall result might appear disappointing. Even the relatively simple task of differentiating centrals and satellites is quite difficult, especially if one wants to classify all galaxies. In the latter case, we get at best a completeness and purity of centrals of 93\% and 84\%, respectively, and of satellites 54\% and 74\%, respectively. Many problems have their origin in issues concerning the basic group catalog (e.g., overmerging, fragmentation). And yet these statistics are very good compared with other group catalogs at high redshift in the literature \citep[e.g.,][]{gerke2005,cucciati2010,gerke2012}. So it is presumably just an unpleasant fact of life that the construction of high-quality group catalogs (at least when using only spatial galaxy information) is very difficult and subjected to several limitations. The reason for this is that groups in contrast to huge clusters and single galaxies exhibit by their nature a rather low-density contrast against the general field which makes them difficult to detect and suffer from problems which can hardly be cured (e.g., overlapping groups in redshift space, interlopers in redshift space; see K09 for a discussion of difficulties in detecting groups).

Can we do better? We discussed in this paper the exploitation of high-quality photo-$z$ to compensate for the incompleteness of spec-$z$ galaxies in our sample. It is very unlikely that one could detect new groups using these photo-$z$ that were not detected before with the spec-$z$, since even these high-quality photo-$z$ have an uncertainty of $\delta_z \sim 0.01 (1+z)$ which amounts to several times the extension of a group along the line of sight. However, the photo-$z$ are quite useful in characterizing groups that are already detected. Especially big groups benefit much from the photo-$z$ information, insofar as they improve the estimation of the group centers significantly ($\lesssim 100$ kpc offsets) and prevent mistakes in assigning most massive galaxies to groups. The inclusion of photo-$z$ also allows unbiased estimates of the corrected richness $N_{\rm corr}$ to an accuracy of $\sim\! 30 \%$.

Would a 100\% sampled spectroscopic survey with the same flux limit produce a ``better'' group catalog?  While a higher sampling rate will find more groups of lower average mass, the figures of merit  such as $g_1$, $g_2$, etc.~(see Sect.~\ref{sec:definitions}), will not improve substantially. These are defined relative to the groups that should have been detectable in the survey. This is seen in the small statistical differences between the 10k and the 20k group catalog (see Fig.~\ref{fig2:catalog_statistics}) and also in the differences between the full and the central region of the 20k sample (see Tab.~\ref{tab2:catalog_properties}). We also performed tests with complete flux limited mock samples, which also suggest that the gain in these statistics would only improve a couple of percent. We would find more groups for a given observed richness $N$, but at any richness $N$ the basic problems in detecting groups (e.g., the overlapping of groups in redshift space, low-density groups, interlopers in redshift space, etc.) would remain. This expectation is also shown by a comparison with the FOF group catalog from the highly complete GAMA survey, whose statistics were also obtained by comparison with the Millennium simulation \citep{robotham2011} and are broadly comparable, as far as we can determine from their paper. For example, their reported fraction of 2WM reconstructed mock groups of 77\% is not substantially better than our value of $\sim\!75\%$ in Figure \ref{fig2:different_associations}.  The construction of high-quality group catalogs is presumably subject to limitations that are intrinsic to the nature of groups and not so sensitive to the details of the spectroscopic survey.

\section{Conclusion}\label{sec:conclusions}

In the first part of this paper, we have presented the construction and properties of the zCOSMOS 20k group catalog. The basic catalog was derived by applying an FOF multi-run algorithm, whose parameters were tuned by realistic simulated mock galaxy catalogs, on the $\sim\!16$,500 high quality-spectroscopic redshifts of the 20k zCOSMOS sample.

The catalog contains 1498 groups in total and 192 groups with more than five spectroscopic members. If pairs are excluded, its one-way completeness is as high as 83\%, and its one-way purity 82\% compared to all groups principally observable within the 20k sample. About 75\% of these groups exhibit a 2WM (i.e., a one-to-one correspondence) to real groups. These statistics are robust over essentially the whole range of richness, above three or more members, and across the whole redshift range. The fraction of spectroscopic galaxies that can be associated to a group decreases from about 35\% to 10\% over the redshift range from $z \sim 0$ to $z \sim 1$. A prominent feature of the catalog is that the number of reconstructed groups traces very accurately the number of real groups for all richnesses.

Comparisons of the 20k group catalog with the 24 mock catalogs obtained from the DM Millennium simulation exhibit some similarities, but also some differences. The number of groups in the 20k catalog are well within the error bars of the number of reconstructed mock groups over a broad range of observed richnesses. However, there are too many small groups with $N = 2$--3 and too few large groups with $N \gtrsim 20$ in the zCOSMOS group catalog compared to the mock catalogs. This could be an indication that the $\sigma_8$ of the Millennium simulation is in fact too large compared to the actual universe in agreement with the latest cosmological measurements. The fraction of galaxies in groups for the total catalog shows fair agreement with the mock catalogs except for $z \gtrsim 0.7$ where the fraction is significantly higher for the actual data. On the other hand, particularly at high redshift and for volume-limited samples, there are apparently more galaxies in groups than expected from mock catalogs.

We do detect clear evidence for the growth of cosmic structure over the last seven billion years because the fraction of galaxies that are found in groups (in volume-limited samples) decreases significantly to higher redshifts.

In the second part of this paper, we have developed a scheme for complementing the group population by those galaxies which have no reliable spectroscopic redshift, but only a photometric one. This was achieved by assigning to all photo-$z$ galaxies a mock-calibrated association probability $p$ for being a member of a given group. With the aid of the mock catalogs we studied the fidelity, distribution, and completeness of photo-$z$ galaxies associated to groups and found that the concept works comparably well for the actual data.

Using the flux limited group population and the membership probabilities, we introduced a probability $p_{\rm M}$ for each galaxy to be the most (stellar) massive of a group. We found that, for the actual data as well as for the mock data, most of the groups of any richness have a clear well defined candidate for their most massive galaxy. The fidelity of $p_{\rm M}$, however, depends sensitively on the measurement errors of the stellar mass. Despite this problem, selecting galaxies with $p_{\rm M} \geq 0.7$ yields a success rate of finding the real most massive galaxies in more than 50\% of cases.

As another application of the membership probability, using the mock catalogs we studied ten estimators for locating the spatial centers of the groups, of which four are based only on spec-$z$ information and six on a combination of spec-$z$ and photo-$z$. We found that all estimators typically depend on both spectroscopic group richness $N$ and projected apparent extension of the group. Typically, the higher $N$ and the more extended a group, the more effective is the consideration of the photo-$z$ information. Also weighting the galaxy position by the inverse of their projected Voronoi area is more effective in high-$N$ groups. We found that the combination of weighting galaxies by their inverse Voronoi areas and by their stellar mass is superior than just using one of these weighting schemes alone. We define ``improved centers'' by a combination of these estimators (without using information from stellar mass) which should yield offsets $\lesssim \! 100$ kpc from the deepest point of the potential well for most groups of any richness class and group extension. According to the mock catalogs, by considering stellar mass even smaller offsets are achievable.

The best of the 10 estimators achieves the successful selection of the galaxy at the potential minimum (not to be confused with the most massive galaxy) for at least half of all mock groups with $N \geq 5$, and for 75\% of all groups it yields offsets of less than 20 kpc from the real group center.

Finally, we investigated the question how well we can define galaxy samples of central and satellite galaxies, where the centrals are defined to be the galaxies lying at the minimum of the gravitational potential. In addition to $p_{\rm M}$ we introduced another probability, $p_{\rm MA}$, which includes beside the stellar mass also information from the local density at the position of the galaxy. While for picking the central galaxy of a group $p_{\rm M}$ and $p_{\rm MA}$ work comparably well; they are most powerful when taking in combination. We found that by applying suitable cuts in $p$, $p_{\rm M}$, and $p_{\rm MA}$, we are able to construct fairly pure samples of either centrals or satellites (typically about 60\%-80\% purity depending on the richness of the group).

If we want to classify all galaxies in a binary way as either centrals or satellites, we can compute the completeness as well as purity for either centrals and satellites. We defined a division such that for the total flux-limited sample the completeness and purity of centrals are 89\% and 81\%, respectively, and of satellites 45\% and 62\%, respectively. By constraining this division to the spectroscopic sample, the completeness and purity of the centrals are 93\% and 84\%, respectively, and of satellites 54\% and 74\%, respectively.

This research was supported by the Swiss National Science Foundation, and it is based on observations undertaken at the European Southern Observatory (ESO) Very Large Telescope (VLT) under the Large Program 175.A-0839.

\bibliographystyle{apj3}
\bibliography{apj-jour,bibliography}

\end{document}